\newif\ifsingle
\newif\ifFullVersion
\ifsingle
\documentclass[11pt,draftclsnofoot, onecolumn]{IEEEtran}		
\else		
\documentclass[10pt,final, twocolumn]{IEEEtran}
\fi

\usepackage{times}
\usepackage{amsmath,dsfont}
\usepackage{amssymb,amsthm}
\usepackage{epsfig,verbatim}
\usepackage{subfigure}
\usepackage{setspace}
\usepackage{color}
\usepackage{cite}
\usepackage{epstopdf}
\usepackage{graphics}
\usepackage{accents}
\usepackage{acronym}
\usepackage[bookmarks,colorlinks]{hyperref}
\usepackage{booktabs}
\usepackage{mathtools}
\usepackage{algorithm}
\usepackage{algorithmic}
\usepackage{enumitem}
\UseRawInputEncoding
\usepackage{makecell}

\usepackage[all=normal,paragraphs=tight,floats=normal,mathspacing=normal,wordspacing=tight,charwidths=tight,mathdisplays=normal,leading=normal]{savetrees}
\ifsingle

\else

\fi % ------------------------------------------

\acrodef{adc}[ADC]{analog-to-digital convertor}
\acrodef{aoa}[AOA]{angle of arrival}
\acrodef{coa}[COA]{curvature of arrival}
\acrodef{crb}[CRB]{Cramér Rao Bound}
\acrodef{cs}[CS]{compressed sensing}
\acrodef{csi}[CSI]{channel state information}
\acrodef{dma}[DMA]{dynamic metasurface antenna}
\acrodef{dtft}[DTFT]{discrete-time Fourier transform}
\acrodef{dnn}[DNN]{deep neural network} 
\acrodef{gps}[GPS]{global positioning system} 
\acrodef{map}[MAP]{maximum a-posteriori probability}
\acrodef{snr}[SNR]{signal-to-noise ratio}
\acrodef{sinr}[SINR]{signal-to-interference-and-noise ratio}
\acrodef{bs}[BS]{base station} 
\acrodef{em}[EM]{electromagnetic} 
\acrodef{iot}[IOT]{Interent of Things}
\acrodef{mimo}[MIMO]{multiple-input multiple-output}
\acrodef{mse}[MSE]{mean-squared error}
\acrodef{pdf}[PDF]{probability density function}
\acrodef{rv}[RV]{random variable}
\acrodef{fec}[FEC]{forward error correction}
\acrodef{lti}[LTI]{linear time-invariant}
\acrodef{mle}[MLE]{maximum likelihood estimation}
\acrodef{wss}[WSS]{wide-sense stationary}
\acrodef{psd}[PSD]{power spectral density}
\acrodef{ris}[RIS]{reconfigurable intelligent surface}
\acrodef{ser}[SER]{symbol error rate} 
\acrodef{ber}[BER]{bit error rate} 
\acrodef{sgd}[SGD]{stochastic gradient descent} 
\acrodef{toa}[TOA]{time of arrival}
\acrodef{isi}[ISI]{intersymbol interference}  
\acrodef{awgn}[AWGN]{additive white Gaussian noise} 
\acrodef{ut}[UT]{user terminal} 
\acrodef{mmw}[mmWave]{millimeter wave}
\acrodef{6g}[6G]{sixth generation}

\IEEEoverridecommandlockouts

\title{Near-Field Wideband Localization using TTD-Based Terahertz Extremely Large-Scale Arrays
}
\author{  
	\IEEEauthorblockN{Qianyu Yang,~\IEEEmembership{Member,~IEEE}, Haiyang Zhang,~\IEEEmembership{Member,~IEEE}, Francesco Guidi,~\IEEEmembership{Member,~IEEE}, \\Anna Guerra,~\IEEEmembership{Member,~IEEE},  %Nir Shlezinger,~\IEEEmembership{Senior Member,~IEEE}, 
    Davide Dardari,~\IEEEmembership{Fellow,~IEEE},
Baoyun Wang,~\IEEEmembership{Senior Member,~IEEE}, \\ and Yonina C. Eldar,~\IEEEmembership{Fellow,~IEEE}
	} 
	\thanks{
	Q. Yang is with the School of Information and Software Engineering, East China Jiaotong University, Nanchang 330013, China (e-mail: 3763@ecjtu.edu.cn). 
    B. Wang is with the School of Communication and Information Engineering, Nanjing University of Posts and Telecommunications, Nanjing, China (e-mail:  bywang@njupt.edu.cn).
    H. Zhang is with the School of Communication and Information Engineering, Nanjing University of Posts and Telecommunications. He is also with the National Mobile Communications Research Laboratory, Southeast University, Nanjing 210096, China (e-mail: haiyang.zhang@njupt.edu.cn).
    F. Guidi is with the National Research Council of Italy, IEIIT, Bologna, Italy (e-mail: francesco.guidi@cnr.it). 
	A. Guerra and  D. Dardari are with the Department of Electrical, Electronic, and Information Engineering “Guglielmo Marconi” - DEI-CNIT, University of Bologna, Cesena, Italy (e-mail: \{anna.guerra3,davide.dardari\}@unibo.it).
    Y. C. Eldar is with the Faculty of Math and CS, Weizmann Institute of Science, Rehovot, Israel (e-mail:  yonina.eldar@weizmann.ac.il).
		}

}
\begin{document}
	
	\maketitle
	\pagestyle{plain}
	\thispagestyle{plain}
	%----------------------------------------------------------------------------------------
	%	ABSTRACT
	%----------------------------------------------------------------------------------------
%\FraCmt{In the title I would not consider "Beam Focusing" (we already say "near-field" and it would sound too close to our previous papers) and "Multi-User"}
%\FraCmt{Alternative titles: \\
%T1: Wideband Localization with Near-Field Terahertz Wireless Systems %\\
%T2: Near-Field Wideband Localization with Terahertz Wireless Systems}

\begin{abstract}

%The synergy between extremely large-scale antenna arrays and Terahertz technology in the sixth-generation networks establishes a near-field wideband transmission environment. This form of signaling enables the generation of focused beams, introducing new capabilities for wireless localization.
%\FraCmt{I think we should discuss more the availability of a large BW at THz}
%This paper explores a beam-focusing design to simultaneously localize multiple sources from received wideband signals in the near-field. We propose a direct localization estimation method based on the curvature-of-arrival of spherical wavefronts from wideband signals, where hybrid analog/digital array architectures with true time delayers (TTDs) are introduced to deal with the spatial-wideband effect from wideband transmission.
%\FraCmt{If we account for wideband TX signals, we should anticipate it}
%We establish a closed-form position error bound to evaluate estimation performance limits and design the array coefficients accordingly. Then, we extend this method to simultaneously localize and focus using a sub-optimal iterative technique without requiring  position knowledge.
%The simulation results show that the proposed array configuration design significantly enhances the performance of near-field wideband localization, while the presence of TTDs further augments the processing capacity of the array for multi-carrier signals.

The synergy between extremely large-scale antenna arrays and terahertz technology in sixth-generation networks establishes a near-field wideband transmission environment, enabling the generation of highly focused beams.
To leverage this capability for multi-source localization, we propose a direct localization method based on the curvature-of-arrival of spherical wavefronts for estimating the positions of multiple near-field users from wideband signals. Furthermore, to overcome the spatial-wideband effect, we introduce a hybrid analog/digital array architecture with true-time-delayers (TTDs).
We derive a closed-form position error bound to characterize the fundamental estimation performance and optimize the analog coefficients of array by maximizing the trace of the Fisher information matrix to minimize this bound. Furthermore, we extend this method to a sub-optimal iterative method that jointly optimizes beam focusing and localization, without requiring prior knowledge of the source positions for array design.
Simulation results show that the proposed array configuration design significantly enhances the performance of near-field wideband localization, while the presence of TTDs effectively mitigates the localization performance degradation caused by spatial-wideband effects.

\end{abstract}

\acresetall	
\bstctlcite{IEEEexample:BSTcontrol}

%----------------------------------------------------------------------------------------
%	Introduction
%----------------------------------------------------------------------------------------

\section{Introduction}

In sixth generation (6G) networks, the integration of extremely large-scale antenna arrays (ELAAs) and the utilization of Terahertz (THz) bands are expected to notably enhance radio frequency (RF) positioning accuracy \cite{6Gcontext}. Specifically, THz wideband radio positioning is expected to undergo significant advancements and widespread adoption, as the wide bandwidth available will result in high spatial resolution, and the feasibility of large antenna arrays enables unprecedented angular resolution \cite{lotti2023radio,chen2022tutorial,sarieddeen2020next}. The transition towards ELAAs and the incorporation of THz frequency bands entail more than just an increase in bandwidth and the ability to stack more antenna elements on a given platform \cite{jornet2024evolution,ZhaEtAl:J25,petrov2023near}. These advancements involve operating with fewer RF chains than elements and integrating analog processing to a certain extent in 6G networks \cite{ahmed2018survey,mendez2016hybrid,shlezinger2023ai}. They also bring about substantial alterations in the electromagnetic characteristics of the wireless environment, transitioning from planar wave to spherical wave propagation, thereby signal transmission occurrs within the radiating near-field region \cite{6Gcontext,EbaEtAl:J25}. 
%\FraCmt{I think we missed a discussion on Localization with THz wireless systems. See as an example: \cite{lotti2023radio,chen2022tutorial,sarieddeen2020next}. Please refer also to Josep Jornet papers on THz antennas, like \cite{jornet2024evolution,petrov2023near}}

Hybrid architectures, which utilize fewer RF chains than antenna elements, often employ phase shifter (PS) networks \cite{ioushua2019family, lavi2023learn,levy2024rapid} to connect multiple antennas to less RF chains, forming arrays with high-dimensional analog processing and low-dimensional digital processing. 
However, in THz systems, the spatial-wideband effect leads to beam split, causing non-overlapping main lobes of the array gain corresponding to the lowest and highest subcarriers, resulting in a significant degradation of positioning estimation accuracy \cite{Ahmet2023,HanEtAl:J24}. Specifically, PSs are typically optimized for a single frequency (i.e., the central frequency), while in wideband systems, different frequencies across the band experience different phase shifts due to their varying wavelengths. Since phase shifters do not account for these variations across frequencies, the direction of the main beam "splits" or shifts off the intended target as the frequency moves away from the center. This shift creates an unintended offset in the beam’s direction, especially noticeable in systems with a large frequency range.
%\AnnCmt{I think that we can better explain the concept of beam split. Here we can say that beam split occurs because phase shifters, when used alone, introduce phase adjustments that are typically optimized for a single frequency (i.e., the central frequency). In wideband systems, however, different frequencies across the band experience different phase shifts due to their varying wavelengths. Since phase shifters do not account for these variations across frequencies, the direction of the main beam "splits" or shifts off the intended target as the frequency moves away from the center. This shift creates an unintended offset in the beam’s direction, especially noticeable in systems with a large frequency range.}
%\FraCmt{As I said before, we need to better describe the literature on THz localization, otherwise it is not clear which is the benefit of the proposed beamfocusing design} 
While matrix projection techniques aim to alleviate this issue \cite{widehybrid}, their effectiveness is hampered by the frequency-independent nature of PSs \cite{Ahmet2024}.

To address this challenge, true-time delayers (TTDs) can be integrated into arrays to enable frequency-dependent beamforming. 
Specifically, the time delay introduced by TTDs induces frequency-dependent phase shifts across different subcarriers, thereby enabling distinct beamforming patterns to be applied to each subcarrier signal with different frequency and effectively mitigating the beam squint effect.
Moreover, to manage the high power consumption and hardware complexity of TTDs, mainstream TTD-based hybrid arrays typically incorporate a limited number of TTDs between the PSs and RF chains \cite{TTDs3,TTDs2,TTDs,widefocus}. 
%several architectures leveraging TTDs have been proposed for wideband systems \cite{TTDs3,TTDs2,TTDs,widefocus}.
%\FraCmt{I have question: the impact of the bandwidth is linked also with the carrier freq. (I mean the ratio $f_c/BW$) In other words: the impact of $BW=1$GHz of bandwidth at $f_c=4$GHz is much more evident than the impact at $f_c=1000$ GHz. Can we discuss a little bit more this point in the paper?maybe also in the next section} 
For instance, \cite{TTDs3} analyzed the array gain loss resulting from the beam splitting effect in THz massive MIMO systems and addressed the issue through the incorporation of TTDs. Meanwhile, \cite{TTDs2} extended TTD-based wideband beamforming to the near-field scenario. Based on the conventional fully-connected TTD-based architecture, \cite{TTDs} considered practical maximum delay constraints of TTDs and proposed a hybrid serial–parallel TTD configuration that relaxes the maximum delay requirements per TTD. Furthermore, \cite{widefocus} introduced a partially-connected TTD-based hybrid architecture aimed at improving energy efficiency and reducing the number of TTDs required.
%such as fully-connected hybrid architectures \cite{TTDs,TTDs2} and partial-connected hybrid architectures \cite{widefocus}. 
These studies demonstrate that these configurations can achieve near-optimal performance in terms of communication rates while minimizing power consumption. Nevertheless, the utilization of TTD-based hybrid architectures for wideband localization in the radiating near-field has been relatively overlooked in existing literature.  

Differently from far-field localization scenarios \cite{AghEtAl:J25,ErcEtAl:J25}, in the radiating near-field, where wireless systems employing ELAAs at high frequency bands are expected to operate, the  spherical shape of the wavefront becomes non-negligible~\cite{bibtex15}. Such wavefronts introduce new degrees of freedom that can be leveraged to enhance wireless localization, enabling direct and precise localization \cite{9335528,abu2021near,bibtex2,10017173,wang2024near,DarDecGueGui:J22,gasst2025dcd}. This method involves estimating positions, including distance and azimuth of users, directly from the curvature of the spherical wavefront, known as the curvature-of-arrival (COA)
%\AnnCmt{Here we can say that the curvature-of-arrival information is encapsulated in the signal phases. Is this still true using wideband signals? or can we extrapolate the ranging information from signal's delays?} 
based localization technique \cite{CRB, bibtex11, bibtex13}.
Recent studies in the realm of 6G have considered COA-based localization, focusing on various aspects such as characterizing the Cramer-Rao bound (CRB) \cite{CRB}, exploring RIS-assisted configurations \cite{bibtex11}, and tracking methods \cite{bibtex13}. Additionally, \cite{singlelocalize,DMAlocalize} inquired near-field localization with hybrid arrays, showcasing the design of analog coefficients in hybrid arrays for near-field localization can achieve a form of beamfocusing. Similar findings were reported for integrated sensing and communications in \cite{nguyen2024joint,cai2024low}. 

Most existing works of near-field wireless positioning focus on narrowband scenarios. However, with the advent of 6G technologies like ELAAs and THz technology systems, many applications operate in wideband near-field environments \cite{widefocus}, which is far less studied compared to its narrowband counterparts. In particular, \cite{WNFisac} investigated near-field processing for THz systems with a sparse transceiver array, but impact of the beam split was overlooked.
%\FraCmt{I suggest to check in that paper if the authors make any consideration about the relation between the central frequency and the bandwidth} 
The work \cite{WNFlocalize} employed a noise subspace correction approach to address the beam split effect, yet it fails to mitigate the array gain loss stemming from this phenomenon. This motivates exploring  wideband near-field multi-user localization using a TTD-based hybrid array. 

In this work, we investigate RF positioning based on maximum likelihood (ML) estimation using wideband signals in the radiating near-field with a hybrid antenna array.
%We extend the narrowband framework proposed for  joint localization and beamfoucing framework  in \cite{DMAlocalize} to the wideband scenarios.
%\FraCmt{I would articulate more this sentence, otherwise it might sound as a minor variation on our previous paper.} 
To mitigate the beam split effect inherent in wideband systems, we employ hybrid arrays incorporating TTD units. Furthermore, we optimize the analog coefficients (i.e., the phase of PSs and the delay of TTDs) to enhance localization performance. 
Based on a novel problem formulation, we derive a closed-form expression for the CRB on ML position estimation. This derivation elucidates how the design of analog coefficients improves localization accuracy in wideband settings by maximizing the Fisher information matrix (FIM). 
Subsequently, we develop an iterative algorithm utilizing alternating optimization to simultaneously achieve beam focusing and user localization.
%We employ TTD-based hybrid arrays to counteract the beam split problem caused by wideband systems. From a mathematical perspective, the introduction of TTDs decouples the optimization of the analog coefficients from the objective capturing the RF localization problem. Based on our novel formulation, we derive a closed-form expression of the CRB in estimating the position, illustrating how our analog coefficients design enhances localization performance in wideband scenarios, and develop an iterative algorithm based on alternating optimization to simultaneously focus the beam while localizing the users. 

%TODO NIR CONTINIUE FROM HERE
Our main contributions are summarized as follows:
\begin{itemize}
    \item { TTD-Based Wideband COA-Based Localization:} We present a COA-based wideband localization method, designed to achieve simultaneous localization and beam focusing in the radiating near field with a TTDs-based hybrid array. 
    \item { Fundamental Localization Limits Analysis:
    %\FraCmt{If we decide to keep titles in the bullet points, I'd say "Fundamental Localization Limits Analysis"}
    } We derive the CRB for the position estimates obtained via the proposed localization algorithm, which demonstrates the connection between the localization performance and the analog coefficients design. Subsequently, we formulate an optimization problem for the design of analog coefficients to minimize the CRB, enhancing the effectiveness of the localization algorithm.
    \item { Beamfocusing Algorithm:} We employ an alternating optimization approach to reframe the analog coefficients coupling optimization problem into two separate sub-problems: one concerning the PSs coefficients and the other the TTDs coefficients. These sub-problems are solved individually based on assumed user positions, allowing us to iteratively converge on an approximate solution to the original problem. Subsequently, we expand our design to include a sub-optimal iterative method that eliminates the need for prior knowledge of user positions.
    \item { Extensive Simulation Analysis:} We conduct a comprehensive numerical analysis to showcase that the proposed solution notably boosts the performance of wideband near-field localization.
\end{itemize}

%The rest of the paper is organized as follows: Section~\ref{sec:Model} describes the considering wideband signal model and antenna architecture. Then Section~\ref{sec:local} introduces the COA-based localization algorithm, analyzes the associated PEB of estimation based on the CRB, and formulates the optimization problem for designing analog coefficients. The proposed optimization algorithm for addressing this issue is expounded in Section~\ref{sec:Solution}. Subsequently, Section~\ref{sec:Sims} showcases the obtained results. Finally Section~\ref{sec:Conclusions} encapsulates the concluding remarks.
The rest of the paper is organized as follows: Section~\ref{sec:Model} describes the considered wideband signal model and antenna architecture. Section~\ref{sec:LocalizationFixed} formulates COA-based localization for a given beampattern, and analyzes its associated CRB.  In Section~\ref{sec:Solution}, we formulate joint beamfocusing and localization as an optimization problem, based on which the proposed  algorithm  is derived.
%\FraCmt{What Section II and III should be rephrased. Too long and sounds too overlapped.} 
Subsequently, Section~\ref{sec:Sims} reports the  numerical results. Finally Section~\ref{sec:Conclusions} provides concluding remarks.

\emph{Notations}: Scalar variables, vectors, and matrices are represented with lower letters, lower bold letters, and capital bold letters, respectively (e.g., $x$, $\bf x$, and $\bf X$, respectively). The term $\mathbb{C}^{N \times N}$ denotes a complex space of dimension ${N \times N}$, the superscripts $({\cdot})^{ T}$ and $({\cdot})^{ H}$ denote the transpose and Hermitian transpose, respectively, $|\cdot|$ is the absolute value operator, $\Vert \cdot\Vert$ is the Frobenius norm, $\text{blkdiag} \left( {\bf X}_1, \cdots, {\bf X}_N \right)$ denotes a block diagonal matrix with diagonal blocks ${\bf X}_1, \cdots, {\bf X}_N$, and $\left[ {\bf x} \right]_n$ denotes the $n$-th elements of vector ${\bf x}$.

%----------------------------------------------------------------------------------------
%	System Model
%----------------------------------------------------------------------------------------

\section{System Model}
\label{sec:Model}

In this section we present the considered wideband near-field localization system model. We first formulate the corresponding received signal model in Section \ref{subsec:receive}, and then introduce the considered array architectures in Section \ref{subsec:array}.

\subsection{Received Signal Model}
\label{subsec:receive}

We consider a multi-user localization scenario in which a base station receives pilot signals from $K$ single-antenna users
%\AnnCmt{I would specify single-antenna users} 
to estimate their locations. The base station comprises a uniform linear array (ULA) with $N$ antenna elements separated by a distance $\Delta$.
%\FraCmt{Can we justify better why the base station is equipped with a ULA and not a uniform planar array (URA), as typically done in 5G and 6G networks? As an alternative, can we consider a URA which is represented in a vectorized way? \textcolor{blue}{Response: since we consider a 2D localization, the ULA is enough, which is also a common configuration for TTD based hybrid array literature like \cite{TTDs,widefocus}.}} 
In the uplink transmission, all users share the pilot spectrum, transmitting $M$ Orthogonal Frequency-Division Multiplexing (OFDM) subcarriers within a bandwidth $B$ at central carrier frequency $f_{\rm c}$. 
Accordingly, the frequency of the $m$-th subcarrier with $m \in {1, \cdots, M}$, defined as $f_{ m}$, is obtained from a uniform division of the frequency interval $ [f_{\rm c}-B/2,f_{\rm c}+B/2]$. When communication takes place in traditional bands (e.g., sub 6 GHz), it is possible to uniformly process each subcarrier based on $f_{\rm c}$. However, in highly wideband (e.g., Thz) systems, the large bandwidth results in the beam splitting effect, and the frequency difference of each subcarrier cannot be ignored.
%\FraCmt{Can we clarify the impact of the bandwidth?}  

To formulate the resulting signal model, let
%\FraCmt{I would move this last part (from "Let...") before eq. \eqref{receive} and then we describe the meaning of each element.} 
${\bf x}_m=\left[ { x}_{m,1}, \cdots,  { x}_{m,n}, \cdots, { x}_{m,N} \right]^T \in \mathbb{C}^{ N}$ denote the received signal components of the $m$-th subcarrier at the base stations's array. Each source emits a standardized pilot signal with unit power. The signal received by the $n$-th antenna of the ULA at a specific discrete time instance can be expressed as:
\begin{align}
{ x}_{m,n} =\sum_{k=1}^{K} x_{m,k,n} +z_{m,n}.  
\label{receive}
\end{align}
In \eqref{receive}, $z_{m,n}$ is the additive white Gaussian noise at the $n$-th antenna for the $m$-th pilot symbol which has variance $\sigma_m^2$,
%\FraCmt{Is it the noise PSD or noise power/variance? Typically $\sigma^2$ indicates the noise variance/power in Watt, whereas the power spectral density is in Watt/Hz and it is indicated with $N_0$.} 
and $x_{m,k,n}$ is the received signal of the $m$-th subcarrier from the $k$-th user at the corresponding antenna. The latter can be expressed as 
\begin{align}
\begin{aligned}
x_{m,k,n} & = {\rm g}_{m,k,n} \,\, c_k, 
\label{each receive}
\end{aligned}
\end{align}
where $c_k$ is the data symbol corresponding to the $k$-th user, which satisfies $ \mathbb{E} \{|c_k|^2\} = 1$, and
${\rm g}_{m,k,n} = \alpha_{m,k}  e^{-j v_{m,k,n}}$ 
represents the channel component, with $\alpha_{m,k}$ and $v_{m,k,n}$ denoting the channel gain coefficient and the phase due to the distance traveled by the signal, respectively.
%\FraCmt{We express the signal for only one frequency, e.g., $f_m$. In this way, doesn't it result in a narrowband representation of a wideband signal? Shouldn't be there the integration over the frequency band? \textcolor{blue}{Response: To the OFDM communication, each subcarrier could be separated using an IDFT, whether it is wideband or narrowband. The The difference is that in narrowband systems, all subcarriers can be approximately as the central frequency $f_{\rm c}$ to process simultaneously, while in wideband system, it is necessary to processed separately for each subcarrier $f_m$ to avoid beam splitting. That is why we introduce the TTD, since the phase shifter is frequency independent that can only provide a unified phase shift for all subcarriers.}} 
The gain is expressed as $\alpha_{m,k}=\frac{c}{ 4\pi f_{ m} d_{k}}$ 
%\AnnCmt{I would add a footnote stating that we neglect the near-field model of amplitudes. Is $\alpha_{m,k}$ a gain or an amplitude? If it is the latter, I would include a square root.} 
with $ d_{k}$ indicating the distance between the array and the user\footnote{Here we neglect the near-field model of amplitudes, as above  variation in channel gain between different antennas of the array is small enough  \cite{Demir2021}.}, and the phase is given by 
\begin{equation} 
\begin{aligned}
\label{phase1}
v_{m,k,n} \triangleq  2 \pi  f_{ m} \frac{ d_{k,n}}{ c}.
\end{aligned}
\end{equation}
In \eqref{phase1}, $ d_{k,n}$ is the distance between the corresponding antenna and the user, and $ c$ is the speed of light.
Consequently, ${\bf x}_m$ can be written as 
\begin{equation} 
\begin{aligned}
\label{x}
{\bf x}_m &=\sum_{k=1}^{K} {\bf g}_{m,k} \, c_k + {\bf z}_m = {\bf G}_m \, {\bf c} + {\bf z}_m,
\end{aligned}
\end{equation}
where $ {\bf g}_{m,k}=\left[ {\rm g}_{m,k,1}, \cdots,  {\rm g}_{m,k,n}, \cdots, {\rm g}_{m,k,N} \right]^T \in \mathbb{C}^{ N \times 1}$ and $ {\bf z}_m=\left[z_{m,1}, \cdots, z_{m,n}, \cdots, z_{m,N} \right]^T \in \mathbb{C}^{N \times 1}$ denote the channel vector and noise vector, respectively. The matrix ${\bf G}_m=\left [ {\bf g}_{m,1}, \cdots, {\bf g}_{m,K} \right ] \in \mathbb{C}^{ N \times K}$ is the channel matrix, and ${\bf c} = \left[ c_1, \cdots, c_K \right]^T \in \mathbb{C}^{ K \times 1}$ denotes the symbol vector.
%\FraCmt{I have probably missed something. Where does the fact that the system is wideband enters? Here the received signal model is the same as per narrowband signal.}

\begin{figure*}
\centering
\includegraphics[scale=0.6]{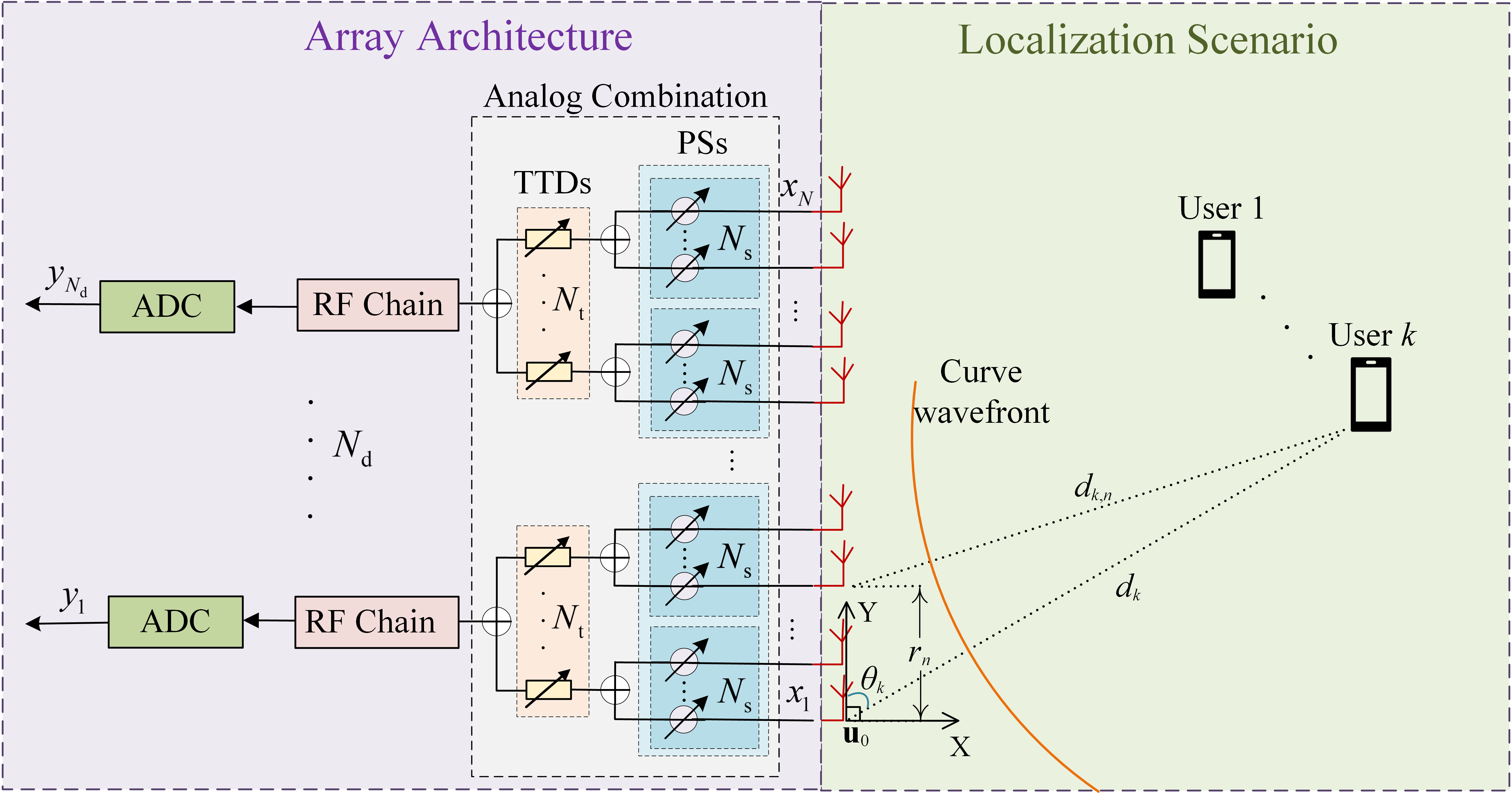}
\caption{System model diagram. Left diagram: the array architectures; Right diagram: the localization scenario. 
%where ${\bf u}_0$ \textcolor{blue}{Anna Comment: check this notation} denotes the reference point.
}
%\vspace{-0.4cm}
\label{fig1}
\end{figure*}

\subsection{Array Architecture} 
\label{subsec:array}
The antenna array utilized by the base station is based on a hybrid architecture, being a widely adopted technique for realizing large scale antennas at high frequencies in an affordable manner~\cite{shlezinger2023ai}.
We specifically consider TTD-based hybrid array configurations, in which large number of antenna elements is connected to a smaller number of RF chains through dedicated analog circuitry comprised of TTD units, as depicted in Fig.~\ref{fig1}. In this architecture, TTD-based preprocessing of the received signal achieves frequency-independent phase adjustment. This TTD-based hybrid array setup employs $N_{\rm d}$ RF chains to connect to $N$ antennas utilizing a restricted level of analog signal processing, each antenna connects to an RF chain via an individual PS and a shared TTD \cite{widefocus}. 

Specifically, the array is partitioned into several linear sub-arrays, each comprising $N_{\rm s}$ elements arranged uniformly. Each linear sub-array is managed by a fully interconnected PS network with a TTD. Subsequently, the TTDs are divided into $N_{\rm d}$ groups, each containing $N_{\rm t}$ TTDs, with each group linked to an RF chain. Consequently, the array incorporates $N_{\rm t}N_{\rm s}$ TTDs and $N$ PSs, satisfying the condition $N = N_{\rm d}N_{\rm t}N_{\rm s}$.
%\AnnCmt{I would add references of real-world implementation/design of this array architecture at Terahertz.}
The signal of the $m$-th subcarrier ${\bf y}_m = \left[ {y}_{ m,1}, \cdots, y_{m,i}, \cdots, {y}_{ m,N_{\rm d}} \right]^T \in \mathbb{C}^{ N_{\rm d} \times 1}$, which is processed in digital after  analog combination, is thus given by
\begin{equation} 
\label{y_h}
{\bf y}_m= {\bf Q}_m \, {\bf x}_m, 
\end{equation}
where ${\bf Q}_m$ is the analog combination matrix, which is determined jointly by the TTDs and PSs as 
\begin{equation}
\label{stru_Q}
{\bf Q}_m = {\bf T}_m {\bf A}.
\end{equation}
In \eqref{stru_Q}, ${\bf T}_m \in \mathbb{C}^{ N_{\rm d} \times N_{\rm d}N_{\rm t}}$ and ${\bf A} \in \mathbb{C}^{ N_{\rm d}N_{\rm t} \times N}$ represent the analog processing carried out by the TTDs and PSs, which are given by
\begin{equation}
\label{stru_T}
{\bf T}_m = \text{blkdiag} \left( e^{j 2 \pi  f_{ m} {\bf t}_{1}}, \cdots, e^{j 2 \pi  f_{ m} {\bf t}_{N_{\rm d}}} \right),
\end{equation}
\begin{equation}
\label{stru_A}
{\bf A} = \text{blkdiag} \left( {\bf a}_{1,1}, \cdots, {\bf a}_{i,l}, \cdots, {\bf a}_{N_{\rm d},N_{\rm t}} \right),
\end{equation}  
respectively. In \eqref{stru_T} ${\bf t}_{i} = \left[ { t}_{i, 1}, \cdots, { t}_{i, N_{\rm t}} \right] \in \mathbb{C}^{ N_{\rm t}}$ is the time delay vector of TTDs at the $i$-th RF chain, with ${ t}_{i, l} \in [0,t_{\text{max}}]$, $i\in \{ 1, \cdots, N_{\rm d} \}$, $l\in \{ 1, \cdots, N_{\rm t} \}$, here $t_{\text{max}}$ being the maximum delay of each TTD. In \eqref{stru_A}, ${\bf a}_{i,l} \in \mathbb{C}^{ N_{\rm s}}$ is the PS coefficient vector connect to the $l$-th TTD at the $i$-th RF chain, its internal element $a_n$ denotes the coefficient of the PS corresponding to the $n$-th antenna with $n \in \mathcal{N}_{i,l} = \{ ((i-1)N_{\rm t}+l-1)N_{\rm s}+1,\cdots,((i-1)N_{\rm t}+l)N_{\rm s}\}$, which is frequency independent and given by
\begin{align}
\label{a_n}
&{ a}_{n} \in \mathcal{F} \triangleq\left\{e^{j \phi_{n}} \mid \phi_{n} \in[0,2 \pi]\right\}.
\end{align}
Hence, the entries of the digital input \eqref{y_h} can be expressed as
%\FraCmt{I typically don't like to close sections with equations. Please add a connection sentence between eq. (10) and Sec. III.}
\begin{equation} 
\label{y_i}
y_{m,i}= \sum_{l = 1}^{N_{\rm t}} e^{j 2 \pi  f_{ m} { t}_{i, l}} \sum_{n \in \mathcal{N}_{i,l}} a_{n} { x}_{m,n}. 
\end{equation}

Our objective is to estimate user positions based on received wideband signal observations $\{ {\bf y}_m\}_{m=1}^M$. As evident from \eqref{y_h} and \eqref{y_i}, the considered array architecture facilitates distinct analog combinations for each received signal ${\bf y}_m$. Therefore, by adjusting the combinations constrained by \eqref{stru_T} and \eqref{stru_A}, the received signal will be subjected to frequency dependent modulation to avoid beam splitting.
%In the next section, we will systematically analyze how this feature influences the near-field wideband localization performance.

%\AnnCmt{Should it be $e^{-j 2 \pi  f_{ m} { t}_{i, l}}$ or $e^{+j 2 \pi  f_{ m} { t}_{i, l}}$?} 
%\AnnCmt{Do we need here to specify that $\phi_n$ is a function of the central frequency only?}
%\AnnCmt{Probably we can anticipate here the definition of $\mathbf{Q}_m$} 

%\FraCmt{Again, I would start from vectors and then I'd define what there is inside each vector element.}

%\section{Localization Algorithm and Error Analysis} \label{sec:local}

\section{Localization with Fixed Analog Combination} \label{sec:LocalizationFixed}
The array architecture modeling in \eqref{y_i} indicates that one can affect the signal processed in digital (and used for localization) by manipulating the analog configuration ${\bf Q}_m$, represented by ${\bf T}_m$ and ${\bf A}$.
To understand how the setting of these analog combining parameters affect the ability to localize multiple users based on wideband transmission,  we start by studying localization for a fixed analog combiner and analyze the associated CRB of estimation rusult. To that aim, we first characterize a suitable localization algorithm based on the maximum likelihood rule before the  CRB analysis. 

%we can derive various outputs from the array. Our goal is to estimate the position of users from the received signal with the designed analog configuration. 

\subsection{Maximum Likelihood Localization}
\subsubsection{Localization Setting}
The utilization of the radiating near-field in communication can be leveraged to enhance localization through the COA estimation of the impinging signal at different antennas. The localization scenario under consideration, depicted in Fig.~\ref{fig1}, establishes a reference point (an antenna) ${\bf u}_0$ situated at $(0,0)$ within the receiving array. The ULA is oriented along the $Y$-axis, with the $k$-th user positioned at $({\rm x}_k,{\rm y}_k)$, a distance of ${d}_k$, and an azimuth angle $\theta_k$ relative to the source (refer to Fig.\ref{fig1}). 
Moreover, the $n$-th antenna of the array is positioned at $\left( 0,{r}_{n}\right)$ with respect to the array's reference point, where ${r}_{n}=(N-1)\Delta$. The specific geometric relationships between coordinates and reference positions can be described as:
\begin{equation}  
\label{positionFormulation}
{\rm x}_k = d_k \cos{\theta_k}, \quad
{\rm y}_k = d_k \sin{\theta_k}.  
\end{equation}
According to the triangular relationship, the distance of the $k$th user from the $n$th antenna can be expressed as \cite{ bibtex13}
\begin{equation}
\label{loca1}
{ d}_{k,n}\left(d_k,\theta_k \right)=\sqrt{{ r}_{n}^{2}+{ d}_{k}^{2}-2 { r}_{n } { d}_{k} \cos \theta_k}.
\end{equation}
%\AnnCmt{I think ${ r}_{n}$ is not defined.}

\subsubsection{Maximum Likelihood Formulation}
The formulation \eqref{loca1} establishes the intricate link between the user's location and the phase distribution, preserving the nonlinearity within the aforementioned output. Hence, one viable approach for user position estimation involves the computation of the maximum likelihood estimation (MLE). By consolidating the received signals across $M$ sub-frequencies, we have ${\bf y} = \left[ {\bf y}_1^{T}, \cdots, {\bf y}_M^{T }\right]^T$, which we express as
\begin{equation} 
\label{y}
{\bf y}= {\bf Q}{\bf G}~\hat{\bf c} + {\bf Q}~{\bf z}, 
\end{equation}
where $\hat{\bf c}={\bf l}_M \otimes {\bf c} \in \mathbb{C}^{MK \times 1}$, ${\bf z} = \left[ {\bf z}_1^{T}, \cdots, {\bf z}_M^{T }\right]^T$, ${\bf Q} = \text{blkdiag} \left( {\bf Q}_1, \cdots, {\bf Q}_M \right)$, ${\bf G} = \text{blkdiag} \left( {\bf G}_1, \cdots, {\bf G}_M \right)$. According to \eqref{stru_T} and \eqref{stru_A}, we have ${\bf Q}_m{\bf Q}_{m}^T = N_{\rm t}N_{\rm s}{\bf I}_{N_{\rm d}}, \forall m$.

We define the unknown parameters vector to be estimated from \eqref{y} as $\boldsymbol{\eta}  = \left[ {\bf d}, \boldsymbol{\theta} \right]\in \mathbb{C}^{2K}$, with ${\bf d} = [d_1,\cdots,d_K]$ and $\boldsymbol{\theta}=[\theta_1,\cdots,\theta_K]$ refer to the possible user positions set in polar coordinates, respectively. The position  MLE  is given by
\begin{equation}
\label{que1}
\hat{\boldsymbol{\eta}} =\underset{\boldsymbol{\eta} }{\arg \max } \log p\left({\bf y} ; \boldsymbol{\eta}, {\bf Q} \right),
\end{equation}
where $ \log p\left({\bf y} ; \boldsymbol{\eta}, {\bf Q} \right)$ is the log-likelihood function of $\bf y$. 

The MLE process is based on an observation time window consisting of $L$ samples. The $l$-th sample is denoted as ${\bf y}(l)$ for $l = 1,\cdots,L$. Given \eqref{y}, once a sufficient quantity of receiving samples is obtained, the log-likelihood function of ${\bf y}$ can be articulated as \cite{ bibtex9}
\begin{align} 
\log p\left({\bf y} ; \boldsymbol{\eta}, {\bf Q} \right) &= \log p\left({\bf y}_1,\cdots, {\bf y}_M; \boldsymbol{\eta}, {\bf Q} \right) \notag\\
&\propto \sum_{l = 1}^{L} \sum_{m = 1}^{M} \Vert {\mathcal{P}}\left[{\bf S}_m \left( \boldsymbol{\eta}, {\bf Q} \right)\right] {\bf y}_m(l) \Vert^{2} \notag\\
&\propto \sum_{m = 1}^{M} \operatorname{tr} \left [ {\mathcal{P}}\left[{\bf S}_m \left( \boldsymbol{\eta}, {\bf Q} \right)\right] {\bf R}_m  \right ],
\label{logp-d}
\end{align}
%\AnnCmt{It would be nice, if possible, to express the log-likelihood as a function of TDDs and PSs matrices (as per eq. 6-7) - I think they both enter in matrix ${\bf S}_m$ through $\mathbf{Q}_m$ and also in ${\bf R}_m$ through ${\bf y}_m$.}
where ${\bf R}_m$ is the empirical average of ${\bf y}_m{\bf y}_m^H $ over the observed time window, and ${\mathcal{P}}$ denotes the projection operator, which, when applied to a matrix ${\bf X}$, is given by
\begin{equation*}
{\mathcal{P}}\left[ {\bf X}\right] =  {\bf X} \left[  {\bf X}^{H} {\bf X}\right]^{-1}  {\bf X}^{H}.
\end{equation*}
Projection in \eqref{logp-d} is applied to ${\bf S}_m \left( \boldsymbol{\eta}, {\bf Q} \right) = {\bf Q}_m{\bf S}_{m}^{\rm d} \left( \boldsymbol{\eta} \right)$, and ${\bf S}_{m}^{\rm d} \left( \boldsymbol{\eta} \right) = \left[ {\bf s}_{m,1}^{\rm d}(\boldsymbol{\eta}_1), \cdots, {\bf s}_{m,K}^{\rm d}(\boldsymbol{\eta}_K) \right] \in \mathbb{C}^{N \times M}$ is the steering matrix of the array with
\begin{equation}
\label{steering}
{\bf s }_{m,k}^{\rm d}(\boldsymbol{\eta}_k)= {\alpha}_{m,k} \left[ e^{-j v_{m,k,1}(\boldsymbol{\eta}_k) }, \cdots, e^{-j v_{m,k,N }(\boldsymbol{\eta}_k) } \right]^T,
\end{equation} 
where $\boldsymbol{\eta}_k = [d_k, \theta_k]$, and $v_{m,i,l} (\boldsymbol{\eta}_k)$ is obtained from \eqref{phase1} and \eqref{loca1} thus determined by $\boldsymbol{\eta}_k$. 

\subsubsection{MLE Computation via Alternating Projection}
Problem \eqref{logp-d} can be solved by the alternating projection (AP) method to avoid the complexity of the multi-dimensional maximization problem \cite{DMAlocalize}. Specifically, we define the vector $\boldsymbol{\eta}_{-k}  = \left[ {\bf d}_{-k}, \boldsymbol{\theta}_{-k} \right]\in \mathbb{C}^{2K-2}$, where ${\bf d}_{-k} = [d_1,\cdots,d_{k-1},d_{k+1},\cdots,d_K]$ and $\boldsymbol{\theta}_{-k} = [\theta_1,\cdots,\theta_{k-1},\theta_{k+1},\cdots,\theta_K]$, then the projection operator ${\mathcal{P}}\left[{\bf S}_m \left( \boldsymbol{\eta}, {\bf Q} \right)\right]$ can be rewritten as 
\begin{equation}
\label{proj_f}
{\mathcal{P}}\left[{\bf S}_m \left( \boldsymbol{\eta}, {\bf Q} \right)\right] \!=\! {\mathcal{P}}\left[{\bf S}_m \left( \boldsymbol{\eta}_{-k}, {\bf Q} \right)\right] + {\mathcal{P}}\left[\overline{\bf s}_m \left( \boldsymbol{\eta}_{k},{\bf Q} \right)\right],
\end{equation}
where ${\bf S}_m \left( \boldsymbol{\eta}_{-k}, {\bf Q} \right)\in \mathbb{C}^{N_{\rm d} \times K-1}$ is the steering matrix based on $\boldsymbol{\eta}_{-k} $ and ${\bf Q}$, and $\overline{\bf s}_m \left( \boldsymbol{\eta}_{k}, {\bf Q} \right)$ is given by
\begin{equation}
\label{proj-f2}
\overline{\bf s}_m \left( \boldsymbol{\eta}_{k}, {\bf Q} \right) = \left( {\bf I}_ {N_{\rm d}} - {\mathcal{P}}\left[{\bf S} \left( \boldsymbol{\eta}_{-k}, {\bf Q} \right)\right] \right) {\bf Q}_m~{\bf s }_{m,k}^{\rm d}(\boldsymbol{\eta}_k),
\end{equation} 
which denotes the residual of ${\bf Q}_m{\bf s }_{m,k}^{\rm d}$ when projected on ${\bf S}_m \left( \boldsymbol{\eta}_{-k}, {\bf Q} \right)$

From \eqref{proj_f}, for a given analog configuration ${\bf Q}$, once $\boldsymbol{\eta}_{-k}$ has been fixed, ${\mathcal{P}}\left[{\bf S}_m \left( \boldsymbol{\eta}, {\bf Q} \right)\right]$ becomes solely reliant on $\boldsymbol{\eta}_k$. Consequently, 
%the corresponding term of \eqref{proj_f} can be neglected, and 
the maximization of \eqref{logp-d} can be addressed alternating iteratively with each iteration involving the maximization concerning a single parameter $\boldsymbol{\eta}_k$ while keeping all other parameters constant.
Thus, by substituting \eqref{proj_f} into \eqref{logp-d}, the problem posed in \eqref{que1} can be streamlined into solving $M$ sub-problems, where the $m$-th sub-problem is formulated as:
\begin{equation}
\begin{aligned}
\label{logp_d2}
\boldsymbol{\eta}_k & = \underset{\boldsymbol{\eta}_k}{\arg \max } \log p\left({\bf y} ; \boldsymbol{\eta}_k, {\bf Q} \right) \\ &\propto\sum_{m=1}^{M} \operatorname{tr} \left[ {\mathcal{P}}\left[\overline{\bf s}_m \left( \boldsymbol{\eta}_{k}, {\bf Q} \right)\right] {\bf R}_m \right]
\end{aligned}
\end{equation}

One can calculate \eqref{logp_d2}  when ${\bf Q}$ is held constant, with the detailed process summarized as Algorithm \ref{alg:AP}.
The initial estimation begins with a single-user scenario to determine the position of the first user. Subsequently, the algorithm fixes the existing estimates and progressively incorporates additional users until reaching the final estimation. This incremental approach aids in refining the localization process step by step.
We note that the proposed Algorithm \ref{alg:AP} is also applicable to narrowband system. In narrowband system case, 
%the TTDs would no longer be necessary, i.e., 
all subcarriers would be tuned by a unified phase shift from PSs without TTDs, that ${\bf Q}$ would be given by ${\bf Q} = \text{blkdiag} \left( {\bf Q}_0, \cdots, {\bf Q}_0 \right)$, where ${\bf Q}_0$ is obtained from \eqref{stru_Q} with $t_{\text{max}}=0$. 

%\AnnCmt{Is Algorithm 1 designed specifically for wideband localization, or is it applicable to narrowband localization as well? I believe that if we disregard the matrix $\mathbf{T}_m$ in $\mathbf{Q}_m$ in Equation 7, we can apply the same algorithmic steps to the narrowband case. If this is the case, then I would prefer to keep it general.}
%
\begin{algorithm}
\caption{AP-based wideband near-field localization.
}
\label{alg:AP}
\begin{algorithmic}[1] %每行显示行号
%\renewcommand{\algorithmicrequire}{\textbf{Initialize:}} 
%\REQUIRE Position mark $\left(d_{\mathcal{M}}, \theta_{\mathcal{M}}, \gamma_{\mathcal{M}} \right)$
\renewcommand{\algorithmicrequire}{\textbf{Initialize:}} 
\REQUIRE Set $t=1$; The maximal number of iteration $ K_{\rm t}$; $\boldsymbol{\eta} \in \mathbb{R}^{2K}$, and ${\mathcal{P}}\left[{\bf S}_m \left( \boldsymbol{\eta}, {\bf Q} \right)\right]= {\bf 0},\forall m$.
\renewcommand{\algorithmicrequire}{\textbf{For} $k = 1:K$} \REQUIRE 
\STATE For all $m$, calculate: \\
$\overline{\bf s}_m \left( \boldsymbol{\eta}_{k}, {\bf Q} \right) = \left( {\bf I}_ {N_{\rm d}} - {\mathcal{P}}\left[{\bf S}_m \left( \boldsymbol{\eta}, {\bf Q} \right)\right] \right) {\bf Q}_m~{\bf s }_{m,k}^{\rm d}(\boldsymbol{\eta}_{k})$.
\STATE Obtain $\boldsymbol{\eta}_{k}^0$ by solving \eqref{logp_d2} with calculated $\overline{\bf s}_m \left( \boldsymbol{\eta}_{k}, {\bf Q} \right)$.
\STATE ${\mathcal{P}}\left[{\bf S}_m \left( \boldsymbol{\eta}, {\bf Q} \right)\right] = {\mathcal{P}}\left[{\bf S}_m \left( \boldsymbol{\eta}, {\bf Q} \right)\right] + {\mathcal{P}}\left[\overline{\bf s}_m \left( \boldsymbol{\eta}_{k}, {\bf Q} \right)\right] $.
\STATE Initialize position set: $[\boldsymbol{\eta}]_k = [\boldsymbol{\eta}_k^0]_1 $, $[\boldsymbol{\eta}]_{K+k} = [\boldsymbol{\eta}_k^0]_2$.
\renewcommand{\algorithmicrequire}{\textbf{End}} \REQUIRE 

\renewcommand{\algorithmicrequire}{\textbf{While} {$t \leq  K_{\rm t}$} \textbf{do}} \REQUIRE
\renewcommand{\algorithmicrequire}{~~\textbf{For} $k = 1:K$} \REQUIRE  
    \STATE Obtain ${\mathcal{P}}\left[\overline{\bf s}_m \left( \boldsymbol{\eta}_{k}, {\bf Q} \right)\right]$ from $\overline{\bf s}_m \left( \boldsymbol{\eta}_{k}, {\bf Q} \right)$ as \eqref{proj-f2}, $\forall m$.
    \STATE Obtain $\boldsymbol{\eta}_{k}^t$ by solving \eqref{logp_d2} with calculated $\overline{\bf s}_m \left( \boldsymbol{\eta}_{k}, {\bf Q} \right)$.
    \STATE Update position set: $[\boldsymbol{\eta}]_k = [\boldsymbol{\eta}_k^t]_1 $, $[\boldsymbol{\eta}]_{K+k} = [\boldsymbol{\eta}_k^t]_2 $. 
    \STATE Update ${\mathcal{P}}\left[{\bf S}_m \left( \boldsymbol{\eta}, {\bf Q} \right)\right] = {\mathcal{P}}\left[{\bf S}_m \left( \boldsymbol{\eta}, {\bf Q} \right)\right] + {\mathcal{P}}\left[\overline{\bf s}_m \left( \boldsymbol{\eta}_{k}, {\bf Q} \right)\right] $.
\renewcommand{\algorithmicrequire}{~~\textbf{End}} \REQUIRE 
    \STATE $t=t+1$.
\renewcommand{\algorithmicrequire}{\textbf{End while}} \REQUIRE 
\renewcommand{\algorithmicrequire}{\textbf{Output:}} \REQUIRE The estimated unknown parameters $\boldsymbol{\eta}$.
\end{algorithmic}
\end{algorithm}
%
%Notably, this algorithm incorporates an initial position estimation phase, as an accurate initialization is paramount for ensuring the overall convergence of the iterative process.

The computation of Algorithm \ref{alg:AP} requires a given ${\bf Q}$. Therefore, the fact that the accuracy of this estimate depends on ${\bf Q}$ allows us to use the above MLE as guidelines for the analog design along with the localization task. 
%For this purpose, a metric is essential for evaluating the positioning error of MLE and, in this regard, in the next we consider the CRB.\FraCmt{How can we initially compute the CRB if we have no idea about where the users are located?}

\subsection{ Cramer-Rao Bound Analysis} 
\label{subsec:CRB}
The CRB serves as a fundamental lower limit on the variance of any unbiased estimator \cite{Van}. The CRB expression derived provides valuable insights into how system parameters and configurations, such as the analog combination, influence the performance of position estimation. %\AnnCmt{You can add a textbook reference, like Van Trees and Kay.}
For the likelihood model in  \eqref{logp-d}, the array output ${\bf y}$ can be regarded as a distribution with the mean vector $\boldsymbol{\mu}$ and covariance matrix $\boldsymbol{\Xi} = \text{blkdiag}(\sigma_1^2 N_{\rm t}N_{\rm s}{\bf I}_{MN_{\rm d}},\cdots,\sigma_M^2 N_{\rm t}N_{\rm s}{\bf I}_{MN_{\rm d}})$, where
\begin{equation} 
\label{mu}
\boldsymbol{\mu}= {\bf Q}{\bf S}^{\rm d}(\boldsymbol{\eta}, {\bf Q})~\hat{\bf c}, 
\end{equation}
and ${\bf S}^{\rm d}(\boldsymbol{\eta}, {\bf Q}) = \text{blkdiag} \left( {\bf S}_1^{\rm d}(\boldsymbol{\eta}, {\bf Q}), \cdots, {\bf S}_M^{\rm d}(\boldsymbol{\eta}, {\bf Q}) \right)$.

%We define the unknown parameters to be estimated as the vector $\boldsymbol{\eta}  = \left[ {\bf d}, \boldsymbol{\theta} \right]\in \mathbb{C}^{2K}$.\AnnCmt{This formulation, i.e., $\boldsymbol{\eta}$, can be used also in the MLE problem for consistency. }
Let $\bf F$ denotes the Fisher information matrix (FIM) for estimating parameter $\boldsymbol{\eta}$ from the data. The elements of the FIM are given by \cite{DOAcrb}
\begin{align} 
[{\bf F}]_{i,j} & =  2\mathcal{R}\left\{\frac{\partial\boldsymbol{\mu}^{H}}{\partial[\boldsymbol{\eta}]_{i}}\boldsymbol{\Xi}^{-1}\frac{\partial\boldsymbol{\mu}}{\partial[\boldsymbol{\eta}]_{j}}\right\}+\operatorname{tr}\left(\boldsymbol{\Xi}^{-1}\frac{\partial\boldsymbol{\Xi}}{\partial[\boldsymbol{\eta}]_{i}} \boldsymbol{\Xi}^{-1}\frac{\partial\boldsymbol{\Xi}}{\partial[\boldsymbol{\eta}]_{j}}\right)\notag \\
& = 2\mathcal{R}\left\{\frac{\partial\boldsymbol{\mu}^{H}}{\partial[\boldsymbol{\eta}]_{i}} \boldsymbol{\Xi}^{-1}\frac{\partial\boldsymbol{\mu}} {\partial[\boldsymbol{\eta}]_{j}}\right\},
\label{fim-ij}
\end{align}
where $[\boldsymbol{\eta}]_{i}$ denotes the $i$-th element of $\boldsymbol{\eta}$.
According to \eqref{mu} and the definition of $\bf Q$, we have the first-order derivative
\begin{equation}
\begin{aligned}
\label{derd}
\frac{\partial\boldsymbol{\mu}}{\partial{\bf d}} &=  {\bf Q} \frac{\partial{\bf S}^{\rm d}(\boldsymbol{\eta}, {\bf Q})}{\partial {\bf d}}~\hat{\bf c}\\
&= \left[ {\bf C} {\bf D}_1^T {\bf Q}_1^T, \cdots, {\bf C} {\bf D}_M^T {\bf Q}_M^T \right]^T,
\end{aligned}
\end{equation}
\begin{equation}
\begin{aligned}
\label{dertheta}
\frac{\partial\boldsymbol{\mu}}{\partial\boldsymbol{\theta}} &= {\bf Q} \frac{\partial{\bf S}^{\rm d}(\boldsymbol{\eta}, {\bf Q})}{\partial \boldsymbol{\theta}}~\hat{\bf c} \\
&= \left[ {\bf C} {\bf B}_1^T {\bf Q}_1^T, \cdots, {\bf C} {\bf B}_M^T {\bf Q}_M^T \right]^T.
\end{aligned}
\end{equation}
Here ${\bf C} = \text{diag}({\bf c})$, ${\bf D}_m$ and ${\bf B}_m$ are respectively defined as
\begin{equation}
\begin{aligned}
\label{Dm}
{\bf D}_m = \left[ \frac{\partial {\bf s}_{m,1}^{\rm d}(\boldsymbol{\eta}_1)}{\partial d_1}, \cdots, \frac{\partial {\bf s}_{m,K}^{\rm d}(\boldsymbol{\eta}_K)}{\partial d_K} \right] \in \mathbb{C}^{N \times K},
\end{aligned}
\end{equation}
\begin{equation}
\begin{aligned}
\label{Bm}
{\bf B}_m = \left[ \frac{\partial {\bf s}_{m,1}^{\rm d}(\boldsymbol{\eta}_1)}{\partial \theta_1}, \cdots, \frac{\partial {\bf s}_{m,K}^{\rm d}(\boldsymbol{\eta}_K)}{\partial \theta_K} \right] \in \mathbb{C}^{N \times K}.
\end{aligned}
\end{equation}
The elements of the matrix in \eqref{Dm} and \eqref{Bm} are given by
\begin{equation}
\begin{aligned}
\label{Dmkn}
[{\bf D}_m]_{n,k} &= \frac{\partial { s}_{m,k,n}^{\rm d}}{\partial d_k} \\
&= -e^{-j v_{m,k,n}}\left( \frac{c}{ 4\pi f_{ m} d_{k}^2} + \frac{j (d_k-r_n \cos \theta_k) }{2d_k d_{k,n}} \right),
\end{aligned}
\end{equation}
\begin{equation}
\begin{aligned}
\label{Bmkn}
[{\bf B}_m]_{n,k} &= \frac{\partial { s}_{m,k,n}^{\rm d}}{\partial \theta_k} \\
&= -e^{-j v_{m,k,n}} \frac{j r_n \sin \theta_k }{2 d_{k,n}}.~~~~~~~~~~~~~~~~~~~~~~~
\end{aligned}
\end{equation}

Then, according to the division of $\boldsymbol{\eta}$, the FIM can be represented as
\begin{equation}
\begin{aligned}
\label{fim}
{\bf F} = \begin{bmatrix}{\bf F}_{\bf dd} & {\bf F}_{{\bf d}\boldsymbol{\theta}} \\{\bf F}_{\boldsymbol{\theta}{\bf d}} & {\bf F}_{\boldsymbol{\theta\theta}} \end{bmatrix},
\end{aligned}
\end{equation}
where each block matrix in \eqref{fim} can be calculated separately as
\begin{equation}
\begin{aligned}
\label{fim-block1}
{\bf F}_{\bf dd} & =  2\mathcal{R}\left\{\frac{\partial\boldsymbol{\mu}^{H}}{\partial {\bf d}^T}\boldsymbol{\Xi}^{-1}\frac{\partial\boldsymbol{\mu}}{\partial{\bf d}}\right\}\\
& = \frac{2}{N_{\rm t}N_{\rm s}} \sum_{m=1}^M \frac{1}{\sigma_m^2} \mathcal{R}\left\{ {\bf C}^H {\bf D}_m^H {\bf Q}_m^H {\bf Q}_m {\bf D}_m {\bf C} \right\},
\end{aligned}
\end{equation}
\begin{equation}
\begin{aligned}
\label{fim-block2}
{\bf F}_{{\bf d}\boldsymbol{\theta}} & =  2\mathcal{R}\left\{\frac{\partial\boldsymbol{\mu}^{H}}{\partial {\bf d}^T}\boldsymbol{\Xi}^{-1}\frac{\partial\boldsymbol{\mu}}{\partial\boldsymbol{\theta}}\right\}\\
& = \frac{2}{N_{\rm t}N_{\rm s}} \sum_{m=1}^M \frac{1}{\sigma_m^2} \mathcal{R}\left\{ {\bf C}^H {\bf D}_m^H {\bf Q}_m^H {\bf Q}_m {\bf B}_m {\bf C} \right\},
\end{aligned}
\end{equation}
\begin{equation}
\begin{aligned}
\label{fim-block3}
{\bf F}_{\boldsymbol{\theta}{\bf d}} & =  2\mathcal{R}\left\{\frac{\partial\boldsymbol{\mu}^{H}}{\partial \boldsymbol{\theta}^T}\boldsymbol{\Xi}^{-1}\frac{\partial\boldsymbol{\mu}}{\partial {\bf d}} \right\} \\
& = \frac{2}{N_{\rm t}N_{\rm s}} \sum_{m=1}^M \frac{1}{\sigma_m^2} \mathcal{R}\left\{ {\bf C}^H {\bf B}_m^H {\bf Q}_m^H {\bf Q}_m {\bf D}_m {\bf C} \right\},
\end{aligned}
\end{equation}
\begin{equation}
\begin{aligned}
\label{fim-block4}
{\bf F}_{\boldsymbol{\theta}\boldsymbol{\theta}} & =  2\mathcal{R}\left\{\frac{\partial\boldsymbol{\mu}^{H}}{\partial \boldsymbol{\theta}^T}\boldsymbol{\Xi}^{-1}\frac{\partial\boldsymbol{\mu}}{\partial \boldsymbol{\theta}} \right\} \\
& = \frac{2}{N_{\rm t}N_{\rm s}} \sum_{m=1}^M \frac{1}{\sigma_m^2} \mathcal{R}\left\{ {\bf C}^H {\bf B}_m^H {\bf Q}_m^H {\bf Q}_m {\bf B}_m {\bf C} \right\},
\end{aligned}
\end{equation}

To convert the estimation results from polar coordinates to Euclidean coordinates, we define the position parameter vector ${\bf p} = [{\rm x}_1, \cdots, {\rm x}_K, {\rm y}_1, \cdots, {\rm y}_K]$. Its corresponding FIM defined as ${\bf F}_{\rm E}$ is obtained by using the chain rule:
\begin{equation}
\begin{aligned}
\label{fE}
{\bf F}_{\rm E} = {\bf J}{\bf F}{\bf J}^T,
\end{aligned}
\end{equation}
with ${\bf J} = \frac{\partial\boldsymbol{\eta}}{\partial{\bf p}^T} \in \mathbb{C}^{2K \times 2K}$ denoting the Jacobian matrix, which can be computed based on the relationship specified in \eqref{positionFormulation}. The CRB is given by the inverse of ${\bf F}_{\rm E}$, formulated as:
%\FraCmt{Thank you for the provided analysis. Is it possible to write the CRB in a more compact way, highlighting the impact of single parameters? Or is it too complex? \textcolor{blue}{Qianyu: We have tried to derive a more specific expression for CRB by the block matrix inversing, unfortunately, it is still difficult to visually demonstrate the factors that affect localization performance due to the complexity of this expression. }}
%\FraCmt{I think we should provide an expression of the PEB, not just leave the sqrt of the trace of matrix $F_E$. In addition, since we retrieve the PEB to solve the ML approach, and since we still have space in the paper, here I think that it would be nice to try to achieve a closed form expression of the CRB which highlights the contribution of the bandwidth and of the sphericity of the wavefront (e.g., near-field). If it is difficult to retrieve, we could consider some simplified geometries/array configurations/....}
\begin{equation}
\begin{aligned}
\label{peb}
 \operatorname{CRB}= \sqrt{\operatorname{tr}\left({\bf F}_{\rm E}^{-1}\right)}.
\end{aligned}
\end{equation}

\subsection{ Discussion} 
\label{subsec:discussion}
Once $\mathbf{Q}$ is given, the position error bound (PEB) of the estimated position $\mathbf{p}$ can be given by the CRB expression in \eqref{peb}. This relationship motivates further investigation into how analog design influences localization performance. Although \eqref{peb} provides a closed-form expression for the PEB, it does not explicitly reveal the dependence of localization accuracy on analog design. Instead, as \eqref{fE} and \eqref{peb}, the structure of the FIMs in \eqref{fim-block1}-\eqref{fim-block4} explicitly demonstrates which factors affect the localization performance. Specifically, we will explore the factors that affect maximizing the trace of the FIMs, in order to achieve the minimization of CRB.
%Once $\bf Q$ is given, the position error bound of estimation for position $\bf p$ can be given by CRB expression \eqref{peb}, that prompts us to further analyze the impact of analog design on localization performance. Unfortunately, though \eqref{peb} provides a closed form expression, it cannot explicitly indicate the impact of analog design on localization performance.Instead, equations \eqref{fim-block1} to \eqref{fim-block4} elucidate how $\bf Q$ influences the FIM. 

Firstly, each block matrix of the FIM can be decomposed into a superposition of $M$ submatrices corresponding to different subcarrier components. This decomposition implies that wideband transmission with richer spectral resources can enhance localization accuracy by reducing the CRB. Specifically, a larger number of subcarriers increases the overall FIM, thereby lowering the CRB as indicated by \eqref{peb}. 
Furthermore, this formulation establishes an intermediate relationship between $\mathbf{Q}$ and the localization CRB: given the user positions and frequency bands, the FIM is primarily determined by\footnote{Although the FIM also depends on the symbol matrix $\mathbf{C}$ as shown in \eqref{fim-block1}-\eqref{fim-block4}, this work assumes no interaction between users and the base station, and thus treats pilot symbols as random variables without optimizing their design. However, in scenarios involving user-base station coordination, pilot optimization could further improve localization performance.} $\mathbf{Q}$. 
%Consequently, the CRB can be computed from the trace of the inverse FIM, and minimizing the CRB can be approximated by maximizing the FIM.
%Here, each FIM block matrix can be decomposed into the superposition of $M$ matrices representing different subcarrier components, which means that richer spectrum resources in wideband transmission will be beneficial for the reduction of CRB, since more subcarrier components leading to higher value of FIM which results lower CRB as \eqref{peb}. Moreover, it provides a mediate contacts between $\bf Q$ and CRB of localization, i.e., for given positions and frequency bands, FIM is determined by $\bf Q$\footnote{Though as equations \eqref{fim-block1} to \eqref{fim-block4}, FIM is also determined by the symbol matrix $\bf C$, this paper does not assume interaction between users and the base station, and therefore does not consider the design of pilots for users (pilot symbols are assumed to be Gaussian random variables).  However, it should be noted that in scenarios where interaction exists, it is feasible to further improve localization performance through pilot design.}, then CRB is obtained by calculating the trace of the inverse of FIM, and it can be found that the minimize problem for CRB can obtain an approximate suboptimal solution by maximizing the FIM.  

Further analysis reveals that the optimal analog design for maximizing different subcarrier components of FIMs varies, highlighting the significance of our work. By incorporating TTD units into the analog beamforming structure, frequency dependent phase control can be achieved, enabling more precise optimization of each FIMs component across subcarriers. This leads to a lower CRB and improved localization accuracy. Therefore, the number of TTDs is a critical factor influencing the localization performance of the proposed wideband near-field localization scheme, as an increase in the former will be beneficial for more flexible frequency independent phase control.

\section{Analog Combination Coefficient Design}
\label{sec:Solution}
While the localization scenario considered in the previous section focused on fixed analog combining (which dictates the beampattern of the ELAA), we next utilize its formulation to jointly localize and analog combining design generated by the TTD-based ELAA. We thus formulate the optimization problem of analog coefficient design in Section \ref{subsec:problem}.
%\FraCmt{I think that we should say something about why Terahertz (e.g., we have electrically large arrays, large available bandwidth in the transmitted signals, etc), and which are its peculiarities. Otherwise, we consider THz only in the numerical results section.}
In Section~\ref{subsec:phase-opt}, we optimize the phase of PSs for a fixed time delay configuration, then optimizing the time delay of TTDs for a fixed PSs configuration. Finally, in Section \ref{subsec:alternate}, we extend these designs to also carry out localization, via an iterative algorithm for joint optimization position estimation and analog coefficient design.

\subsection{ Problem Formulation} 
\label{subsec:problem}

%Though \eqref{peb} provides a closed form expression for CRB, it cannot explicitly indicate the impact of analog design on localization performance.Instead, equations \eqref{fim-block1} to \eqref{fim-block4} elucidate how $\bf Q$ influences the FIM. Here, each FIM block matrix can be decomposed into the superposition of $M$ matrices representing different subcarrier components, which means that richer spectrum resources in wideband transmission will be beneficial for the reduction of CRB. Moreover, it provide a mediate contacts between $\bf Q$ and CRB of localization.
%\FraCmt{The acronym PEB should be defined before.}\FraCmt{Actually, such equations show the impact on the FIM elements and not on the PEB (in between there is also a change of coordinates). But of course, given those expressions, one can imagine the impact of Q on PEB.} 
%Notably, the received signals from the hybrid array are intricately linked with the analog combination matrix $\bf Q$, underscoring the impact of $\bf Q$ on the accuracy of the MLE. 
According to \eqref{stru_Q}, the design of PSs and TTDs contribute to effective localization leveraging the hybrid array setup, where PSs are frequency independent while TTDs are frequency dependent. By optimizing the configurations of PSs and TTDs, the localization process can be enhanced, ultimately leading to improved accuracy and reliability in position estimation.

As discussed in Section~\ref{subsec:discussion}, %instead of minimizing the CRB directly (by minimizing the inverse of the FIM), 
we design $\bf Q$ to maximize the average of the trace of FIM\footnote{The randomness inherent in pilot symbols results in variations in the FIM obtained from each sampling instance. Consequently, it is necessary to compute the statistical average to mitigate the effects of this variability.}, which is expressed as
\begin{equation}
\label{max_pro}
\begin{aligned}
&\max_{\{{\bf Q}_m\}_{m=1}^M} ~ \operatorname{tr}[{\bf F} ] \\
&~~~~~\text{s.t.}~~ \eqref{a_n},\eqref{stru_T}, \eqref{stru_A}, \\
&~~~~~~~~~~ { t}_{i, l} \in [0,t_{\text{max}}]. 
\end{aligned}
\end{equation} 

%Since ${\bf Q}_m$ is determined by ${\bf A}$ and ${\bf T}_m$, in next section, we provide the design scheme for ${\bf A}$ and ${\bf T}_m$, respectively, to obtain the optimal ${\bf Q}_m$ design.
According to \eqref{fim}-\eqref{fim-block4}, by substituting the definition of ${\bf Q}$ and removing the irrelevant variable, problem \eqref{max_pro} can be rewritten as
\begin{equation}
\label{max_pro2}
\begin{aligned}
&\max_{\{{\bf T}_m\}_{m=1}^M,{\bf A}} ~ \sum_{m=1}^M  \operatorname{tr} \left[ {\bf T}_m{\bf A}{\bf U}_m {\bf U}_m^H {\bf A}^H {\bf T}_m^H\right]  \\
&~~\text{s.t.}~~ \eqref{a_n}, \eqref{stru_T}, \eqref{stru_A},\\
&~~~~~~~ { t}_{i, l} \in [0,t_{\text{max}}]. 
\end{aligned}
\end{equation} 
Where ${\bf U}_m = {\bf D}_m + {\bf B}_m$.
Given the structural constraints outlined in \eqref{stru_T} and \eqref{stru_A}, Problem \eqref{max_pro2} is non-convex, primarily due to the coupled nature of the optimization variables, i.e., ${\bf T}_m$ and $\bf A$. To address this challenge, an alternate optimization strategy is employed to disentangle ${\bf T}_m$ and $\bf A$. This iterative alternate approach continues until convergence is achieved, facilitating the optimization process in a computationally feasible manner to approximate the optimal solution.

\subsection{Alternating Optimization for TTDs and PSs} 
\label{subsec:phase-opt}

\subsubsection{ Phase design for given time delay}

For given ${\bf T}_m$, the constrains on $t_{i,l}$ and structure constraint \eqref{stru_T} in problem \eqref{max_pro2} can be removed. Define ${\bf u}_{m,k}$ as the $k$-th column of ${\bf U}_m$ and $\overline{\bf a} = \operatorname{vec} \left( {\bf A} \right) \in \mathbb{C}^{N_{d}N \times 1}$. Then ${\bf T}_m{\bf A}{\bf u}_{m,k} = \left( {\bf u}_{m,k} \right)^T \otimes {\bf T}_m \overline{\bf a}$. Define ${\bf W}_{m,k} = \left( {\bf u}_{m,k} \right)^T \otimes {\bf T}_m$. We can then remove the structure constraint \eqref{stru_A} in \eqref{max_pro2} and rewrite the problem as
\begin{equation}
\label{max_pro2.1}
\begin{aligned}
&\max _{\overline{\bf a}} ~ \overline{\bf a}^H \left( \sum_{m=1}^M \sum_{k=1}^K {\bf W}_{m,k}^H {\bf W}_{m,k} \right) \overline{\bf a}  \\ 
&~~\text{s.t.}~~  \eqref{a_n}.
\end{aligned}
\end{equation}
Given that the zero elements of $\overline{\bf a}$ does not impact the objective function of \eqref{max_pro2.1}, \eqref{max_pro2.1} can be simplified as
\begin{equation}
\label{max_pro2.2}
\begin{aligned}
&\max _{\tilde{\bf a}} ~ \tilde{\bf a}^H \left( \sum_{m=1}^M \sum_{k=1}^K \tilde{\bf W}_{m,k}^H \tilde{\bf W}_{m,k} \right) \tilde{\bf a}  \\ 
&~~\text{s.t.}~~  \eqref{a_n}.
\end{aligned}
\end{equation}
where $\tilde{\bf a} = [a_1,\cdots,a_N]^T \in \mathbb{C}^N$ is the modified version of $\overline{\bf a}$ obtained by removing all the zero elements of $\overline{\bf a}$, and $\tilde{\bf W}_{m,k}$ is the modified version of ${\bf W}_{m,k}$ by removing the elements having the same index as the zero elements of $\overline{\bf a}$.

According to the constraint \eqref{a_n}, solving \eqref{max_pro2.2} can be viewed as a search process for R with the searching space is a product of $N$ complex circles, which is a Riemannian submanifold of $\mathbb{C}^N$.
%The phase constraint in \eqref{a_n} maintains the non-convex nature of problem \eqref{max_pro2.2}. However, solving \eqref{max_pro2.2} can be viewed as a search process to maximize the objective function of \eqref{max_pro2.2} by determining $\{{a_n}\}_{n=1}^N$.
%\AnnCmt{Check this notation} 
%The search space forms a product of $N$ complex circles, constituting a Riemannian submanifold of $\mathbb{C}^N$. 
Consequently, addressing \eqref{max_pro2.2} can be accomplished using the Riemannian conjugate gradient (RCG) algorithm.
Specifically, the solution to \eqref{max_pro2.2} is iteratively updated according to 
\begin{equation}
\label{b_update}
\begin{aligned}
[\tilde{\bf a}^{(p+1)}]_n = \frac{[\tilde{\bf a}^{(p)} + {\boldsymbol \epsilon}^{(p)} {\boldsymbol \xi}^{(p)}]_n}{\vert [\tilde{\bf a}^{(p)} + {\boldsymbol \epsilon}^{(p)} {\boldsymbol \xi}^{(p)}]_n \vert}, \forall n.
\end{aligned}
\end{equation}
Here, $[\tilde{\bf a}^{(p+1)}]_n$ represents the updated point derived from the current point $[\tilde{\bf a}^{(p)}]_n$, with $p$ indicating the iteration count. 

The solution utilizing the RCG method can be seen as an enhancement over the projection-based approach. ${\boldsymbol \epsilon}^{(p)}$ denotes the Armijo step size \cite{bibtex15} in the $p$-th iteration, while ${\boldsymbol \xi}^{(p)}$ signifies the search direction at the point $\tilde{\bf a}^{(p)}$, lying within the tangent space of the complex circle manifold at that specific point. This direction is given by
\begin{equation}
\label{eta}
\begin{aligned}
w{\boldsymbol \xi}^{(p)}& = -\text{grad}~g\left( \tilde{\bf a} \right) + \\
&~~~~\zeta^{(p)} \left( {\boldsymbol \xi}^{(p-1)} - \text{Re} \left({\boldsymbol \xi}^{(p-1)} \circ \tilde{\bf a}^{(p)\dag} \right)\circ \tilde{\bf a}^{(p)} \right),
\end{aligned}
\end{equation}
where $\dag$ and $\circ$ signify conjugation and the Hadamard product respectively, $\zeta^{(p)}$ is selected as the Polak-Ribiere parameter \cite{bibtex15}. $\text{grad}~g\left( \tilde{\bf a} \right)$ represents the Riemannian gradient of $ g\left( \tilde{\bf a} \right)$, computed as the orthogonal projection of the Euclidean gradient of $ g\left( \tilde{\bf a} \right)$, given by
\begin{equation}
\label{grad}
\text{grad}~g\left( \tilde{\bf a} \right) = \nabla g\left( \tilde{\bf a} \right) - \text{Re} \left(\nabla g\left( \tilde{\bf a} \right) \circ \tilde{\bf a}^{(p)\dag} \right)\circ \tilde{\bf a}^{(p)},
\end{equation}
where $\nabla g\left( \tilde{\bf a} \right)$ is the Euclidean gradient of $ g\left( \tilde{\bf a} \right)$, which is defined from \eqref{max_pro2.2} as
%\FraCmt{We should include some comments after the equation below and before discussing the time delay design for a given phase}
\begin{equation}
\label{deta_b}
\nabla g\left( \tilde{\bf a} \right) = 2 \left( \sum_{m=1}^M \sum_{k=1}^K \tilde{\bf W}_{m,k}^H \tilde{\bf W}_{m,k} \right) \tilde{\bf a} .
\end{equation}
With the given time delay setting of TTDs, we can compute $\{{a_n}\}_{n=1}^N$ iteratively with \eqref{b_update} to \eqref{deta_b}, then we consider to determine the time delay with a given phase of PSs.

\subsubsection{Time Delay Design for Given Phase}

For given $\bf A$, \eqref{max_pro2} reduces to 
\begin{equation}
\label{max_pro2.3}
\begin{aligned}
&\max_{\{{\bf T}_m\}_{m=1}^M} ~ \sum_{m=1}^M  \operatorname{tr} \left[ {\bf T}_m{\bf A}{\bf U}_m {\bf U}_m^H{\bf A}^H {\bf T}_m^H\right]  \\
&~~\text{s.t.}~~  \eqref{stru_T},~ { t}_{i, l} \in [0,t_{\text{max}}], ~\forall i,l.
\end{aligned}
\end{equation} 
We define $\overline{\bf v}_m = \operatorname{vec} \left( {\bf T}_m \right)$ and ${\bf Z}_{m,k} = \left( {\bf A}{\bf u}_{m,k} \right)^T \otimes {\bf I}_{N_{\rm d}} $ and rewrite \eqref{max_pro2.3} as
\begin{equation}
\label{max_pro2.4}
\begin{aligned}
&\max _{\tilde{\bf v}_m} ~ \sum_{m=1}^M \tilde{\bf v}_m^H \left( \sum_{k=1}^K \tilde{\bf Z}_{m,k}^H \tilde{\bf Z}_{m,k} \right) \tilde{\bf v}_m  \\ 
&~~\text{s.t.}~~  \tilde{\bf v}_m =  \left[ e^{-j 2 \pi  f_{ m} {\bf t}_{1}}, \cdots, e^{-j 2 \pi  f_{ m} {\bf t}_{N_{\rm d}}} \right]^T, \\
&~~~~~~~ { t}_{i, l} \in [0,t_{\text{max}}], ~\forall i,l.
\end{aligned}
\end{equation}
Here $\tilde{\bf v}_m$ is the modified version of $\overline{\bf v}_m$, and $\tilde{\bf Z}_{m,k}$ is the modified version of ${\bf Z}_{m,k}$ by removing the elements having the same index as the zero elements of $\overline{\bf v}_m$. 
Problem \eqref{max_pro2.4} can be characterized as $N_{\rm d}N_{\rm t}$ single-variable optimization problem concerning ${ t}_{i, l}$. 
Although deriving a closed-form solution for \eqref{max_pro2.4} is complex, the problem can be effectively addressed using numerical techniques like iterative one-dimensional search methods for practical computation.
Specifically, define a searching set $\mathcal{S}=\left\{ 0, t_{\text{max}}/Q, 2t_{\text{max}}/Q,\cdots, t_{\text{max}}\right\}$, where $Q$ is the sample number.
In each iteration, each ${ t}_{i, l}$ is updated by one-dimensional searching in $\mathcal{S}$.

% ------------------- Alternating Localization and Beamfocusing ------------------------

\subsection{Alternating Localization and Array Coefficient Design}
\label{subsec:alternate}

%\AnnCmt{Instead of using the term "Beamfocusing," could we refer to it as "Array Coefficient Design" to align with the title of Section III?} 
The analog design method previously discussed relies on location information that may not be readily accessible in practical scenarios. Nonetheless, drawing inspiration from \cite{DMAlocalize}, an efficient alternating optimization approach can be employed for joint analog domain design and localization.
%\FraCmt{I did not get well what is different from our paper \cite{DMAlocalize} in the alternating localizaiton/beamfocusing design. We should better stress the differences.}
Specifically, \cite{DMAfocusing} reveals that even when the focus position at the receiving array deviates from the actual source position, improvements in the estimated results are observed. This enhancement persists despite discrepancies between the focus position and the user's true location, with the focus gain being present but attenuated. This suggests that once an initial estimation of the source location is acquired, iterative refinement towards the genuine source position is achievable by updating the focus position. Simultaneously, the adjustment of ${\bf Q}_m$ can be synchronized with the iterations in the AP algorithm. In essence, an alternating algorithm can be devised by integrating the update of ${\bf Q}_m$ into Algorithm \ref{alg:AP}, which is summarized as Algorithm~\ref{alg:Alternating}.

\begin{algorithm}
\caption{Alternating localization and beamfocusing}
\label{alg:Alternating}
\begin{algorithmic}[1] %每行显示行号
\renewcommand{\algorithmicrequire}{\textbf{Initialize:}} 
\REQUIRE Initialize the matrix $\{{\bf Q}_m^1 \}_{m=1}^M $ by the definition with random PSs setting ${\bf A}^1$ and TTDs setting ${\bf T}_m^1$ which satisfies $t_{i,l}\in[0,t_{\text{max}}]$; $t=1$;  Iterations limit $K_{\rm t}$. Obtain the initial position estimation as $({\bf d}^1, \boldsymbol{\theta}^1)$ by Algorithm \ref{alg:AP} (Steps 1-4).
\renewcommand{\algorithmicrequire}{\textbf{While} {$t \leq  K_{\rm t}$} \textbf{do}} \REQUIRE  
\renewcommand{\algorithmicrequire}{~~\textbf{For} $m = 1:M$} \REQUIRE 
    \STATE Estimate $(d_k^t, \theta_k^t))$ by maximizing \eqref{logp_d2} with given ${\bf Q}_m^t  $.
    \STATE Update position set: $[{\bf d}]_k = d_k^i $.
\renewcommand{\algorithmicrequire}{~~\textbf{End}} \REQUIRE     
    \STATE Update ${\bf A}^t$ via solving \eqref{max_pro2.1} with ${\bf T}_m= {\bf T}_m^t,\forall m$.
    \STATE Update ${\bf T}_m^t$ via solving \eqref{max_pro2.3} with ${\bf A}= {\bf A}^t$.
    \STATE Update receive vector ${\bf y}^{t+1}$ from ${\bf Q}_m^{t+1}$ by \eqref{y}.
    \STATE $t=t+1$
\renewcommand{\algorithmicrequire}{\textbf{End while}} \REQUIRE
\renewcommand{\algorithmicrequire}{\textbf{Output:}} \REQUIRE The position estimation $({\bf d}, \boldsymbol{\theta}) = ({\bf d}^{K_{\rm t}}, \boldsymbol{\theta}^{K_{\rm t}})$.
\end{algorithmic}
\end{algorithm}

In Algorithm~\ref{alg:Alternating}, we utilize the receiving vector derived from random coefficient responses to initialize the position estimation. Subsequently, we update the coefficients of ${\bf Q}_m$ based on the outcomes of the preceding position estimation to obtain new observations. Different from the alternating algorithm in \cite{DMAlocalize}, in this paper the coefficient update is targeted at the wideband muti carrier scenarios, which require setting analog coefficients for multiple subcarriers separately, and the relevant design methods have been discussed in Section~\ref{subsec:phase-opt}. This iterative procedure, repeated at each receiving time slot, enables a step-by-step convergence of the estimated result towards the true source location.

%The introduction of the $\bf Q$ update in Algorithm~\ref{alg:Alternating} compared to Algorithm~\ref{alg:AP} leads to higher observation time demands, given that each $\bf Q$ update necessitates a re-observation of the signal. However, it is essential to highlight that the utilization of TTD-based hybrid arrays can substantially diminish the need for RF chains. Numerical assessments have demonstrated that this approach not only offers cost advantages related to the array but also notably decreases the requisite signal processing capacity and data processing time. These benefits help offset the extended observation time required in the algorithm.

%-----------------------------------------------------------------------------------%	NUMERICAL EVALUATIONS
%-----------------------------------------------------------------------------------

\section{Numerical Evaluations}
\label{sec:Sims}

\subsection{Simulation Parameters} 
\label{parameter}
In this section, we provide numerical results to demonstrate the efficiency of proposed localization algorithm and analog coefficients design scheme. 
In our experiments, the receiver array is set at configured as line array lying on the Y-axis with the array's reference position situated at $(0,0)$ m, to positioning two users in X-Y plane, i.e., $K = 2$. In this case, the position of two users is set as $(8~{\rm m},\pi/3)$ and $(8~{\rm m},\pi/4)$ in polar coordinates, respectively. For convenience, the noise variance $\sigma_m^2$ is set to adjust SNR of each sub-carrier at antennas as a unified value, i.e., $\text{SNR} = 10 \text{log}(\Vert{\bf x}_m\Vert^2/(N\sigma_m^2)),~\forall m$. Furthermore, the other basic simulation parameters are given in Table~\ref{tab1}, which is exploited throughout our simulations unless otherwise specified.
%\NirCmt{make code publicly available on GitHub and put the link here}
%\FraCmt{Are the parameters in the table taken from somewhere and related to THz OFDM? We have a very large Bandwidth: how many subcarriers?}
%\FraCmt{I think that in the following we should better investigate the impact of the bandwidth, different distances (how moving out the near-field region performance changes), etc. } \AnnCmt{I agree, I would also increase the number of users, or provide at least one result with a larger number of users.}

The numerical indicators of localization performance are represented by the root mean square error (RMSE), calculated as
\begin{equation}
\label{rmse}
\mathsf{RMSE}= \frac{1}{K}\sum_{k=1}^K \sqrt{\frac{1}{N_{\rm c}} \sum_{n_c=1}^{N_{\rm c}}  e_{k,n_c}^2},
\end{equation}
where $N_{\rm c}$ is the number of Monte Carlo iterations, $e_{k,n_c}^2=  |\big( {\rm x}_k, {\rm y}_k \big) - \big( \hat{\rm x}_k^{(n_c)}, \hat{\rm y}_k^{(n_c)} \big)|^{2}$ is the squared localization error (i.e., the distortion) of the $k$-th user, $\big( \hat{\rm x}_k^{(n_c)}, \hat{\rm y}_k^{(n_c)} \big)$ are the estimated coordinates at the $n_c$-th Monte Carlo simulation.
%\AnnCmt{I think we should also average on different users' positions on the plane.}

\begin{table}[t]
\centering
\caption{Experiment Parameters}
\begin{tabular}{|c|c|}
\hline
\textbf{Parameter}                 & \textbf{Value}                             \\ \hline
Central frequency                  & $f_{\rm c} = 300$ GHz                             \\ \hline
Bandwidth                    & $B = 30$ GHz                                    \\ \hline
Subcarrier number                   & $M = 12$                                     \\ \hline
Sample number               & $L = 256$ \\ 
                     \hline
Array antenna number               & $N = 256$ \\ 
                     \hline
RF chain number               & $N_{\rm d} = 8$ \\ 
                     \hline   
TTD number per RF chain               & $N_{\rm t} = 16$ \\ 
                     \hline                      
Antenna spacing                      & $\Delta = 5\times10^{-4}$ m                               \\ \hline
The maximum time delay   & $t_{\rm max} = 5$ ns                               \\ \hline
%User number               & $K = 2$ \\                           \hline  
%User 1 polar coordinates                      & $(8~{\rm m},\pi/3)$                                \\ \hline
%User 2 polar coordinates                      & $(8~{\rm m},\pi/4)$                                \\ \hline
Near-field distance             & $d_F = 32.5$ m                             \\ \hline
\end{tabular}
\vspace{0.4cm}
\label{tab1}
\end{table}

\subsection{Numerical Results}
\label{ssec:results}

\begin{figure}
\centering 
\includegraphics[width=1\columnwidth]{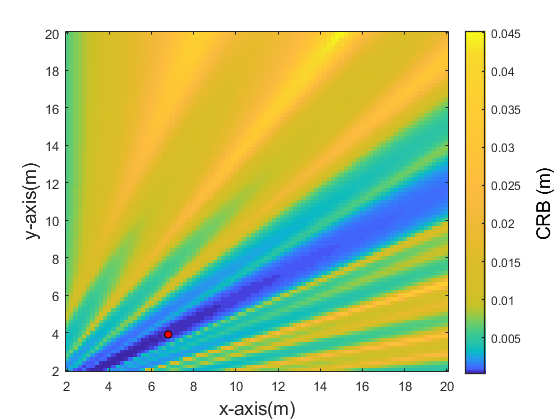}
\caption{The heatmap for the CRB on user position estimate by varying user positions, SNR is fixed as -10 dB. The magenta square denotes the array position. The red point is the predetermined position that analog design focuses on.
}
\vspace{0.4cm}
\label{fig-heat}
\end{figure}

\begin{figure}
\centering 
\includegraphics[width=0.9\columnwidth]{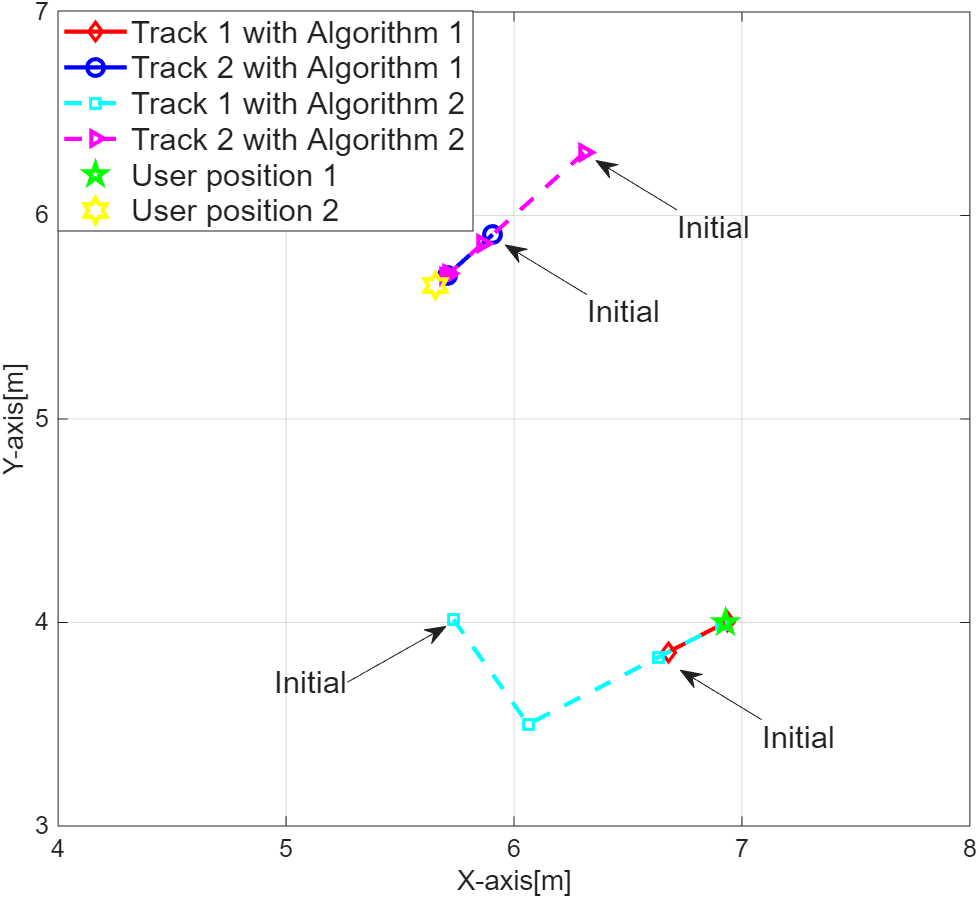}
\caption{
Position estimations track map for each iteration of Algorithm~\ref{alg:AP} and Algorithm~\ref{alg:Alternating} with SNR $\mathsf{SNR} = -10$ dB. 
}
  \vspace{0.4cm}
\label{fig-track}
\end{figure}

\begin{figure}
\centering 
\includegraphics[width=1\columnwidth]{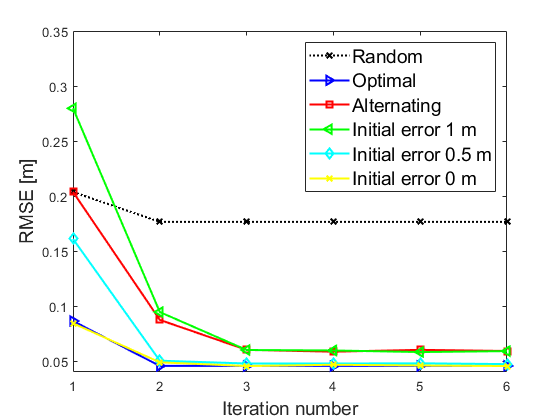}
\caption{Contrast of the number of iterations required for convergence via multiple contrast schemes with $\mathsf{SNR} = -5$ dB.
}
\vspace{0.4cm}
\label{fig2}
\end{figure}

We first demonstrate that the design of $\bf Q$ realizes a form of beam focusing, for this purpose, a heatmap is provided in Fig.~\ref{fig-heat}.
Specifically, we consider a $20 \times 20 , \text{m}^2$ area with a resolution of $0.1$ meters, where the antenna array is positioned at the origin. One user traverses all grid points and stays at each grid point for a positioning cycle to transmit pilot signals, while $\mathbf{Q}$ is designed to focus on the predetermined focal point as shown in Fig.~\ref{fig-heat} (i.e., $\mathbf{Q}$ is derived by solving problem \eqref{max_pro2} based on the FIMs corresponding to the fixed focal point). The array then localizes the user using the actual received pilot signal with this fixed $\mathbf{Q}$ in each positioning cycle.
As illustrated in Fig.~\ref{fig-heat}, the CRB for localization estimation is significantly reduced only when the user's true position is near the focal point of $\mathbf{Q}$. This observation confirms the relationship between the proposed simulated beamforming design and the actual user position, thereby validating our claim that $\mathbf{Q}$ realizes a form of beam focusing in near field wideband localization.
%We consider a $20\times 20 ~ {\rm m}^2$ grid of points with the resolution of $0.1$ meters, with the array located on the origin, and the user transmit the pilot signal on each grid point, then the $\bf Q$ is designed to focus on the fixed point as shown (i.e., the $\bf Q$ is obtained by assuming the user is located on the setting fixed point, then calculate the corresponding FIM to formulate problem \eqref{max_pro2} and solve.). The array would localize the user based on the actual received pilot signal with the fixed $\bf Q$. We can see that the CRB would be significantly improved only when the user position is set near the focus point of $\bf Q$, which proves our view. 

Then, to demonstrate the effectiveness of the proposed localization algorithm, we provide one localization track map in the Monte Carlo simulation, as shown in Fig.~\ref{fig-track}. In Algorithm~\ref{alg:AP}, localization is performed under a given $\mathbf{Q}$, though this setup is not realistic in practice since the true user position is unknown, we assume that $\mathbf{Q}$ is perfectly designed to ensure a fair comparison with Algorithm~\ref{alg:Alternating}. By contrast, Algorithm~\ref{alg:Alternating} alternates between localization and $\mathbf{Q}$ design in an iterative manner. As observed, the estimated position trajectories of both algorithms gradually converge from the initial estimate toward the true user position as iterations proceed. Moreover, it is worth noting that Algorithm~\ref{alg:AP} achieves a more accurate initial position estimate compared to Algorithm~\ref{alg:Alternating} and converges faster to the true position. This clearly highlights the impact of $\mathbf{Q}$ design on the overall localization performance.

To evaluate the efficacy of the analog coefficient design in the joint localization and beamfocusing algorithm, we initially explored the convergence behavior of Algorithm~\ref{alg:Alternating}, as illustrated in Fig.~\ref{fig2}.
%\NirCmt{Can we also include heatmap results like we had in our previous paper to show that beamfocusing is gradually achieved?}
%\FraCmt{At the first iteration, even the random approach gives an error of about 20cm for a SNR of -5dB, which is very good from a physical perspective! So a reviewer might guess what is the advantage of proposing our approach, which requires much more computation. Consider two things: 1. One can say, after first iteration, even with random phases we can refine the position estimate with other solutions. 2. for antenna designers, having random phases would imply a much lower cost in the antenna fabrication. So, if we don't guarantee much better performance, they will still prefer the random solution.} 
Here, the curve ``Alternating" refers to localization using the proposed Algorithm~\ref{alg:Alternating} with a random analog coefficient set as initial. For comparative analysis, we established three contrasting schemes with different initial error as 0 m, 0.5 m and 1 m. Specifically, in above contrast schemes, we still localization with Algorithm~\ref{alg:Alternating}, but assume that we have an indistinct pre knowledge of users locations, where the location error is a Gaussian-distributed random variable centered on the corresponding setting standard deviation, the initial analog coefficient ${\bf Q}_m^1$ in Algorithm~\ref{alg:Alternating} is obtained by solving \eqref{max_pro2.1} and \eqref{max_pro2.3} with such initial locations.
The extreme schemes are also provided, involve localization via Algorithm~\ref{alg:AP} with predetermined analog setting. In such schemes, ``Random" indicates that $\bf A$ and ${\bf T}_m$ are randomly set within specific constraints, while ``Optimal" signifies that $\bf A$ and ${\bf T}_m$ are iteratively optimized by solving \eqref{max_pro2.1} and \eqref{max_pro2.3} with the perfect knowledge of users locations.
%\AnnCmt{In this latter case, is the optimization working with perfect knowledge of users' locations?} 

In Fig.~\ref{fig2}, it is evident from the results that the RMSE of the proposed alternate scheme gradually converges from the level of the random scheme towards close to that of the optimal scheme, while the deviation of the convergence result depends on the initial error. Once the initial error is small enough, e.g., 0.5 m, it can achieve the almost same performance as the optimal scheme which has the perfect knowledge of users locations.
Although the alternate scheme generally necessitates more iterations for convergence compared to the optimal scheme, this is attributed not only to the initial position estimation but also to the additional updates for the analog design. Nonetheless, the iteration process for analog design proves to be efficient, by iteratively analog coefficient design through Algorithm~\ref{alg:Alternating}, the localization accuracy has been improved from decimeter level to centimeter level.
%requiring at most one additional iteration to converge close to the optimal solution. 
%\AnnCmt{Is this true? Do we need proof of convergence? From the figure, it appears that there is a constant bias between the optimal and alternating approaches. Another possibility is to plot the RMSE performance by assuming the position is known with a certain level of uncertainty. This can be done by substituting the true position with a value generated from a Gaussian-distributed random variable centered on the true position, using different standard deviations. In the plot, we can consider various scales for the standard deviation, including the extreme case where it equals zero, which should correspond to the "Proposed Optimal" case.} 

\begin{figure} 
  \centering 
  \subfigure[RMSE versus different SNR.]{ 
    \label{fig3.1}
    \includegraphics[width = 0.95\linewidth]{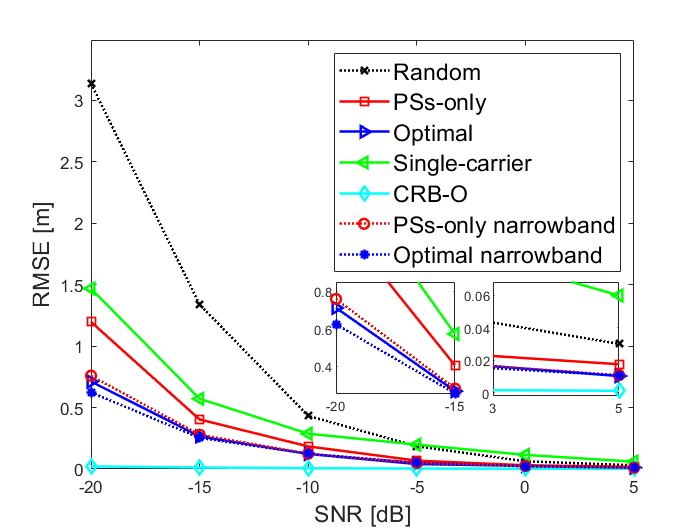} 
  } 
  \subfigure[CRB versus different SNR.  ]{ 
    \label{fig3.2} 
\includegraphics[width = 0.95\linewidth]{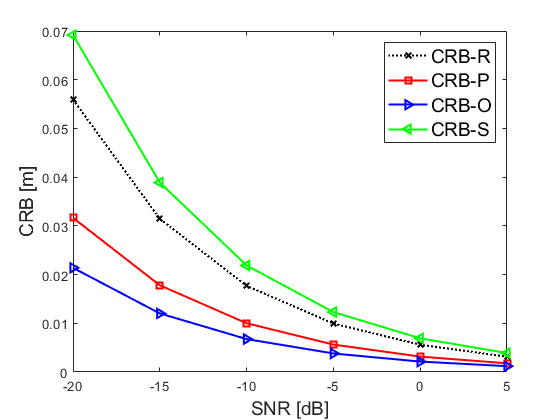} 
  }
\caption{RMSE contrast of multiple contrast schemes under different SNR with array setting. (a) RMSE-SNR; (b) CRB-SNR} 
  \label{fig3} 
  \vspace{0.4cm}
\end{figure}

We illustrate the effectiveness of the analog design scheme we proposed in Fig.~\ref{fig3}, displaying the RMSE of location estimation across various schemes under varying SNRs. As previously discussed, the analog design scheme attains optimal performance when the position information is known. Therefore, we utilize the aforementioned optimal scheme as a benchmark for performance comparison and calculate the CRB under this optimal scenario denoted as ``CRB-O". To showcase the performance enhancement derived from the analog design and further highlight the performance gains resulting from the incorporation of a time delay device, we include the random scheme and the PSs-only scheme for comparison, above analog coefficients of schemes are design randomly, and derived by solving \eqref{max_pro2} with $t_{\rm max} = 0$, serving as an additional reference point in the analysis, respectively. Additional, the single-carrier scheme is also provided to show the performance improvement from the multi-carriers introduction, where the single-carrier scheme is obtained from \cite{DMAlocalize} by set $M=1$. Finally, to evaluate the impact of the bandwidth to optimal scheme and phase-only scheme, we consider a narrowband case, i.e., ``Optimal 2" and ``Phase-only 2", in such case, the the bandwidth is set as 300 MHz.

In Fig.~\ref{fig3.1},
%\FraCmt{It is quite surprising that, for a SNR=-20 dB, the optimal approach allows to attain an error lower than 1m. Is it the SNR per antenna or the overall SNR? \textcolor{blue}{Response: The SNR is the overall SNR as we defined in simulation parameters part. \textcolor{red}{OK, it seems that it accounts for all the antennas} Besides, the good performance in the extreme low SNR that I think comes from the pre knowledge of user positions in ${\bf Q}$ design, i.e., ${\bf Q}$ has focused on the true user location, that assist the system in enhancing the received signal.\textcolor{red}{Does this pre-knowledge of user positions in ${\bf Q}$ design come from the beam focusing after first iteration (even if we say that there is random PS design) or does it come from "outside" (e.g., we assume that there is a prior information about where UEs might be)?.} In Fig.3, the design of initial ${\bf Q}$ come from a prior information about where UEs might be, and to  demonstrate the optimal performance, prior information is set as the true users location (the result of prior information with distortion is shown in Fig.2).}} 
it is evident that the random scheme exhibits the highest RMSE compared to the other schemes in low SNR scenarios. Through the analog design achieved by solving \eqref{max_pro2}, a substantial enhancement in RMSE is observed. Besides, it can be seen that the PSs-only scheme would close to the optimal scheme in the narrowband case, since the beam split effect could be ignored in such case. 
Then in the wideband case, scenarios characterized by high SNRs, the proposed optimal localization scheme closely approaches the performance level of the CRB in comparison to the PSs-only scheme. While in low SNR environments, the former demonstrates around 40\% of the performance enhancement achieved by the latter, it indicates the necessity of the introduction of TTDs for the wideband system. 
An intriguing observation in the single-carrier scheme is that while it exhibited the worest performance in high SNR scenarios, at low SNR levels, its focus obtained through analog design nearly matched the performance of the PSs-only scheme. This futher indicates that the PSs network did not fully exploit the benefits of multi-carrier systems in wideband scenarios, underscoring the importance of incorporating TTDs in multi-carrier settings.
To elucidate the outcomes depicted in Fig.~\ref{fig3.1}, we present the CRB results from various schemes in Fig.~\ref{fig3.2}. Here, ``CRB-R", ``CRB-P" and ``CRB-S" are computed using \eqref{peb} with analog design derived from the random scheme, the phase-only scheme and the single-carrier scheme, respectively, under the bandwidth $B_1$. The comparison of these CRB values mirrors the trends observed in the RMSE performance of their corresponding schemes, suggesting that the performance enhancement exhibited by the proposed scheme stems from its optimization of the CRB.
%\FraCmt{It is quite surprising that, for a SNR=-20 dB, the CRB comprised between 2cm and 7cm. How is the SNR evaluated? \textcolor{blue}{Response: The CRB only provide a theoretical lower limit for estimate, and usually, RMSE can only approach it under high SNR conditions, while the gap increases as the SNR decreases.} \textcolor{red}{Yes, I know that RMSE approaches CRB only for high SNR. But, for low SNR, the CRB typically degrades a lot, showing that under bad noise conditions it is hard to localize a user. Here we instead say that, ideally, the system allows to achieve, in the worst case, an error of 7 cm. I think that somehow in the FIM we inject some (strong) prior information.} \textcolor{blue}{Indeed, the optimal ${\bf Q}$ is designed with prior information, and another improvement I think come from the multi sub-carriers, e.g., for the single-carrier case, if ${\bf Q}$ is random setting, the CRB would increase to about 0.25 m in -20 dB SNR (with the perfect prior information, it would improve to 0.07 m), but in multi-carrier case, i.e, $M=12$, even the random scheme could also archive to 0.06 m.} } 

\begin{figure}
\centering 
\includegraphics[width=0.9\columnwidth]{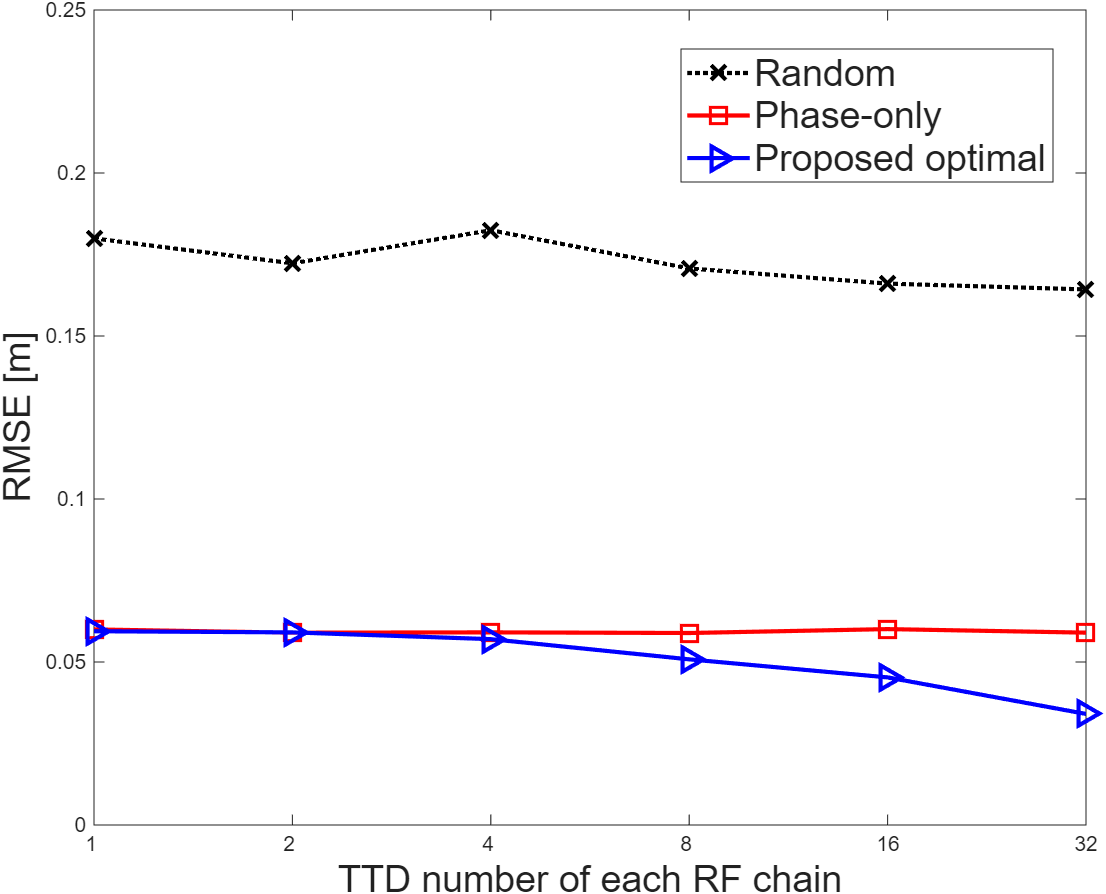}
\caption{RMSE contrast of multiple contrast schemes under different TTDs number of each RF chain with $\mathsf{SNR} = -5$ dB. 
  }
  \vspace{0.4cm}
\label{fig4}
\end{figure}

In Fig.~\ref{fig4},
%\FraCmt{What is distortion contrast?} 
we observe the influence of the number of TTDs connected to each RF chain, denoted as $N_{\rm t}$, on the localization efficiency.
%\FraCmt{Can we say that we consider as an upper limit the case when the number of TDD is the same as that of the antennas?} 
To ensure a fair comparison, we adjust the number of PSs
%\AnnCmt{Phase shifters?} 
connected to each TTD, denoted as $N_{\rm s}$, in accordance with the variation of $N_{\rm t}$ to maintain the condition $N = N_{\rm d}N_{\rm t}N_{\rm s}$ as specified in Tabel~\ref{tab1}, that the upper limit of $N_{\rm t}$ is $N_{\rm t} = N/N_{\rm d}$ with $N_{\rm s} = 1$.
The proposed optimal scheme demonstrates improved performance with increasing $N_{\rm t}$. This enhancement is attributed to the increased processing flexibility across multiple sub-carriers facilitated by the addition of more TTDs, thereby decelerate the beam splitting effect. In contrast, the performance of the random scheme and the phase-only scheme remains unaffected by changes in $N_{\rm t}$. However, this performance enhancement is not as pronounced when $N_{\rm t}$ is less than the number of sub-carriers $M$. In such instances, the performance of the proposed optimal scheme tends to approach that of the phase-only scheme.
Conversely, when the number of TTDs is maximized (i.e., one TTD per antenna), the localization performance exhibits the most substantial improvement. However, deploying a dedicated TTD for each antenna in an ELAA system may be impractical due to prohibitive cost considerations.
%\AnnCmt{In this figure you can add also the curves for a different level of SNR, e.g. -15 dB.}

\begin{figure}
\centering 
\includegraphics[width=1\columnwidth]{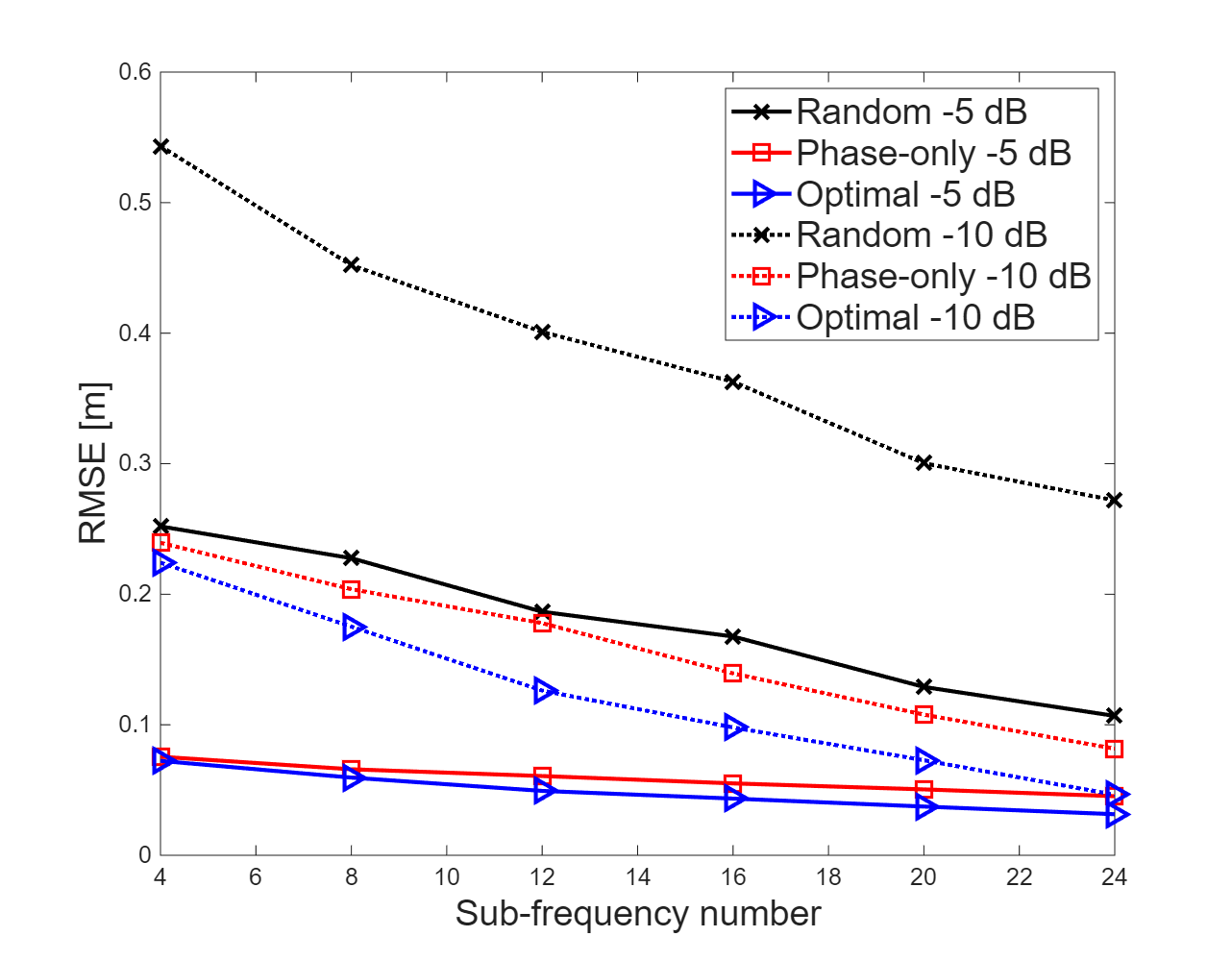}
\caption{RMSE contrast of multiple contrast schemes under different subcarrier number with $\mathsf{SNR} = -5$ or -10 dB. 
  }
  \vspace{0.4cm}
\label{fig5}
\end{figure}

To further elucidate the impact of the relationship between the number of sub-carriers and the number of TTDs connected to each RF chain on the performance of the proposed localization scheme, we present Fig.~\ref{fig5} showcasing the RMSE variations across different sub-carrier numbers. The results reveal that the performance of all schemes improves as the number of sub-carriers increases. This improvement can be attributed to the availability of more received information, a concept supported by the definitions of the FIMs derived in \eqref{fim-block1}-\eqref{fim-block4}. Furthermore, it is evident that in scenarios with a small number of sub-carriers, the phase-only scheme closely matches the performance of the proposed optimal scheme. However, as the number of sub-carriers grows, a performance gap emerges between the two schemes, particularly when the number of sub-carriers exceeds $N_{\rm t}$. This observation underscores that the primary advantage of integrating TTDs lies in resolving the limitation of PSs network in accurately processing multiple sub-carrier signals due to their frequency independence. The effectiveness of TTDs in managing multiple sub-carriers is contingent upon their quantity. A substantial performance enhancement necessitates a greater number of TTDs connected to each RF chain than the number of sub-carriers.

%To further indicate the impact of the relationship between the sub-carrier number and the TTDs number of each RF chain on the performance of proposed localization scheme, we provide Fig.~\ref{fig5} to illustrate the RMSE contrast under different sub-carrier numbers. The result shows that the performance of all schemes increases with the increase of the number of sub-carrier, this is because more sub-carrier mean more receivable information, which can be confirmed by the definition of the FIMs we derived in \eqref{fim-block1}-\eqref{fim-block4}. Moreover, it can be observed that the phase-only scheme achieve a close performance with the proposed optimal scheme in a small sub-carrier number case, while the performance gap would grow with the increase of the sub-carrier number until the sub-carrier number is bigger than $N_{\rm t}$. This indicates that the main advantage of introducing TTDs is to solve the problem of PSs being unable to accurately process multiple sub-carrier signals due to its frequency independence, and the ability of TTDs to handle multiple sub-carriers depends on their number. A significant improvement requires the number of TTDs connected to each RF chain to be greater than the number of sub-carrier.

	%----------------------------------------------------------------------------------------
	%	CONCLUSIONS
	%----------------------------------------------------------------------------------------

\section{Conclusions}
\label{sec:Conclusions}

In this work, our focus delved into wideband near-field multi-user localization utilizing a TTDs-based hybrid array. Initially, we elucidated the models pertaining to radiating near-field position estimation systems within the specified antenna architecture. Subsequently, we derived the CRB for the proposed estimation systems, and accordingly delineated the optimization strategy for the adjustable coefficients of the hybrid array to enhance position estimation accuracy, and introduced an efficient algorithm for simultaneous position estimation and adjustable coefficients design. The numerical outcomes underscored that the meticulous design of analog coefficients within the hybrid array wielded a substantial impact on enhancing near-field multi-source localization performance. Furthermore, the incorporation of TTDs was observed to improve the processing capabilities of the array, particularly in scenarios involving multi-carrier signals.

%In this work, we studied near-field multi-source position estimation based on the RF chain reduction array. We first presented the models for radiating near-field position estimation systems for the three different types of antenna architectures, including fully digital array, hybrid array, and DMA, where the latter two can be expressed in the same model formulation. We then formulated the optimization of the adjustable coefficients of the hybrid array and DMA to improve the accuracy of the position estimate and proposed an efficient algorithm for joint position estimate and adjustable coefficients design. Numerical results demonstrated that the design of adjustable coefficients of RF chain reduction array could significantly improve the near-field multi-source localization performance, reaching a fully digital implementation with at least one order of RF chain reduction.

%\ifFullVersion
%----------------------------------------------------------------------------------------
%	APPENDICES
%----------------------------------------------------------------------------------------

	%----------------------------------------------------------------------------------------
	%	BIBLIOGRAPHY
	%----------------------------------------------------------------------------------------
%	\newpage
	\bibliographystyle{IEEEtran}
	\bibliography{IEEEabrv,refs}

\end{document}